\documentclass[trackchanges]{aastex7}
\usepackage{isotope}
\usepackage{soul} 
\usepackage{hyperref}

\hypersetup{linkcolor=red,citecolor=cyan, filecolor=cyan,urlcolor=magenta}

\newcommand{\ergscmA}{\,erg\,s$^{-1}$\,cm$^{-2}$\,\AA$^{-1}$}
\newcommand{\kms}{\,km\,s$^{-1}$}
\newcommand{\ergs}{\,erg\,s$^{-1}$}
\usepackage{booktabs}   
\usepackage{multirow} 
\usepackage{longtable}
\usepackage{rotating} 
\usepackage{epstopdf}
\shorttitle{V1405 Cas}
\shortauthors{Habtie \& et al.}

\begin{document}
\received{July 22, 2025}
\revised{July 10, 2026}
\accepted{July 20, 2026}
\title{Optical Spectroscopy and Temporal Evolution of the Nova V1405 Cas}

\author[orcid=0000-0001-9827-738X,sname='Habtie']{Gesesew Reta Habtie}
\affiliation{Debre Berhan University, Department of Physics, 445, Ethiopia}
\email[show]{gesesewreta@dbu.edu.et}  

\author[orcid=0000-0002-5440-7186, sname='Das']{Ramkrishna Das} 
\affiliation{SN Bose National Centre for Basic Sciences, Department of Astrophysics and High Energy Physics, 700106, India}
\email{ramkrishna.das@bose.res.in}

\author[orcid=0000-0001-8525-3442, sname='Aydi']{Elias Aydi} 
\affiliation{Department of Physics \& Astronomy, Texas Tech University, Box 41051, Lubbock, TX, 79409-1051, USA}
\email{elias.aydi@gmail.com}

\author[orcid=0000-0001-5541-2836, sname='Dubovsky']{Pavol A. Dubovsky} 
\affiliation{Vihorlat Observatory in Humenne, Mierova 4, Humenne 06601, Slovakia}
\affiliation{Variable Star Section of the Slovak Astronomical Society and the Slovak Union of Astronomers}
\email{var@kozmos.sk}
\begin{abstract} 
This paper presents the findings from our study of nova V1405~Cas over the first 1051 days after the outburst. The study includes an analysis of the photometric light curve evolution, along with a detailed spectroscopic evolution. Photometric analysis shows that the nova is a very slow nova, with a decline timescale of $t_2 \approx 165$ days.  The mass of the white dwarf is calculated as \( M_{\rm WD} \sim 0.7\~M_{\odot} \). The secondary star is a low-mass main-sequence star, with a mass of \( M_{\rm sec} \approx 0.43\~M_{\odot} \). Spectral observations show initial dominance by Balmer emission lines accompanied by prominent P Cygni profiles during the first +339 days. These features disappeared in later epochs, being replaced by high-ionization coronal lines, indicating that the nova had transitioned to the coronal phases by day +371. To investigate the physical conditions of the ejecta and the central source, we conducted photoionization modeling using \textsc{cloudy}. Our model reveals a gradual increase in the temperature and luminosity of the system, and suggests that the ejecta were primarily composed of He, N, Fe, Ne, and Ca, with noticeable temporal variations in their relative abundances. The estimated mean ejected mass is approximately $1.10~\times~10^{-4}\~M_{\odot}$, which is relatively high and suggests a low-mass white dwarf. Our optical spectroscopic and photometric analyses, combined with detailed photoionization modeling, indicate that the white dwarf in Nova V1405 Cas is unlikely to be of the ONeMg type.	
\end{abstract}

\section{Introduction} 
Classical novae (CNe) are thermonuclear explosions that occur on the surfaces of white dwarfs (WDs) in interacting binaries, where hydrogen-rich material is transferred from a companion star through Roche lobe overflow or stellar winds \citep{warner1995, 2008Bodebook, 2021ARA&A..59..391C}. As material accumulates on the WD, rising pressure and temperature eventually ignite nuclear burning. Energy generation proceeds primarily through hydrogen-fusion reactions (pp-chain and CNO cycle), which under degenerate conditions trigger a thermonuclear runaway (TNR) \citep{1998PASPGehrz, 2016PASPStarrfield}. This results in a nova eruption characterized by a dramatic rise in luminosity by factors of $10^3$–$10^6$. Observations show that CNe typically eject $10^{-7}$–$10^{-3}\~M_{\odot}$ of material at velocities of $\sim$ 200–5000 \kms, with an average ejected mass of about $2 \times 10^{-4}\~M_{\odot}$ \citep{2008Bodebook, 2020ApJ...905...62A}.

Nova Cas 2021 (V1405 Cas) was initially spotted as a transient candidate by Yuji Nakamura, registering at 9.6 magnitudes (unfiltered) on 2021-03-18.424 UT (JD 2459291.924)\footnote{\url{http://cbat.eps.harvard.edu/unconf/followups/J23244760+6111140.html}}. Subsequently, it was confirmed to be a CNe by \citet{2021ATel14471Maehara} and \citet{2021ATel14472Taguchi}. The nova was initially captured in the He/N phase \citep{2024MNRASAydi}, and approximately one month later it transitioned into the Fe \textsc{ii} phase \citep{2021ATel14577Shore, 2021ATel14614Munari}. Subsequently, as its luminosity decreased, it reverted to He/N phase spectra again \citep{2021ATel14622Shore}. All novae typically go from He/N to Fe II, and the eventually to He/N \citep{2024MNRASAydi}. 

V1405 Cas has been extensively studied across multiple wavelengths since its eruption on March 18, 2021 \citep{2023arXiv230204656VValisa, 2024BlgAJ..40...13M,  2023ApJTaguchi, 2025NatAs.tmp..249A}. These observations have provided valuable insights into its photometric and spectroscopic evolution. Following the eruption, V1405 Cas exhibited a pre-maximum phase lasting nearly two months \citep{2025NatAs.tmp..249A}. After reaching its primary maximum, the nova displayed multiple short-lived secondary maxima over several months, maintaining a bright plateau before eventually declining into the nebular phase \citep{2023arXiv230204656VValisa, 2023ApJTaguchi}. This behaviour closely resembles that of very slow novae such as HR Del and V723 Cas. Early spectra revealed highly ionized lines, notably He \textsc{ii} and N \textsc{iii}, which were replaced by lower-ionization lines such as N \textsc{ii}, Si \textsc{ii}, and O \textsc{i} between days 10 and 24. This rapid spectral evolution suggests an expanding photosphere \citep{2023ApJTaguchi}.

In 2021, the nova was monitored by the \textit{Fermi Large Area Telescope} (LAT). During the rapid optical brightening episode occurring $\sim$ 55 days after discovery, LAT observations yielded no significant $\gamma$-ray emission, providing only upper limits on the GeV flux \citep{2021ATel14620....1G}. A separate LAT analysis of data obtained $\sim$ 70 days after discovery reported faint $\gamma$-ray emission from a position consistent with V1405~Cas \citep{2021ATel14658....1B}. 
This detection prompted follow-up multi-wavelength observations, including radio monitoring with the Very Large Array (VLA), which clearly detected V1405~Cas at frequencies above 5~GHz $\sim$ 84 days after optical discovery \citep{2021ATel14731Sokolovsky}. Subsequent \textit{Swift}-XRT observations in December 2021 reported a faint super-soft X-ray component around day 270 after discovery, although the low number of detected photons made this identification uncertain \citep{2021ATel15111....1P}.

\citet{2022ATel15796Munari} suggested that V1405 Cas may have occurred on an ONeMg WD based on the appearance of Ne emission lines more than 600 days after the eruption. Similarly, the presence of N~\textsc{III} and Al~\textsc{II} lines in the ejecta has also been interpreted as indicative of an ONe WD \citep{2023ApJTaguchi,2024ATel16876....1P}. However, the identification of an ONeMg progenitor based solely on the presence of Ne emission lines is not definitive. Strong [Ne~\textsc{v}] emission, for example, can arise during the super-soft source (SSS) phase as a result of high-ionization conditions produced by the hot central WD. Furthermore, Al enrichment is not unique to novae occurring on ONeMg WDs and has also been reported in some CO novae (e.g., V3890~Sgr; \citealt{2020ApJ...895...80O}). Consistent with this interpretation, \citet{2024AstL...50..317T} found that the neon and iron abundances are close to their solar values and concluded that the available evidence does not support an ONeMg WD classification. 

In this paper, we present a detailed study of the optical spectral evolution of the nova V1405 Cas following its outburst. In addition, we examine the photometric evolution of the nova. To interpret the observed spectra, we employ the photoionization code \textsc{Cloudy} and construct a simple phenomenological model assuming spherical geometry, allowing us to estimate key physical and chemical parameters of the system during the outburst. Our study spans $\sim$ 1050 days after the outburst, revealing an exceptionally slow evolutionary pattern, placing V1405 Cas among the slowest novae known to date. The structure of the paper is as follows: Section~\ref{sec:observation} describes the observational techniques and data reduction methods; Section~\ref{sec:results} presents the analysis and discussion of the results, including the light curve in Section~\ref{lc}, spectroscopic evolution in Section~\ref{sec3}, and photoionization modeling in Section~\ref{pma}.

\section{Data}\label{sec:observation}
\subsection{Photometric data} \label{phot1}
We used publicly available photometry in the \textit{BVRI} bands from the American Association of Variable Star Observers (AAVSO)\footnote{\url{https://www.aavso.org/}}, which consists of contributions from hundreds of dedicated amateur and professional astronomers. These data are utilized to generate a long-term light curve in the four bands. The optical light curve for the first 1050 days following the outburst is shown in Figure~\ref{fig:lightcurvenher21}.  

\subsection{Spectroscopic data}\label{spec1}
Spectroscopic observations of the object were conducted using the Himalayan Faint Object Spectrograph Camera (HFOSC) mounted on the 2\,m Himalayan Chandra Telescope (HCT), operated by the Indian Institute of Astrophysics (IIA), Bangalore, India; the Low-dispersion, Intermediate-resolution Spectrograph for Astronomy (LISA) attached to the Celestron 11-inch Schmidt--Cassegrain Telescope (C11 SCT), England; the Celestron 11-inch and 14-inch Schmidt--Cassegrain Telescopes (C11 and C14 SCT), Slovakia; and the PlaneWave 20-inch Corrected Dall--Kirkham Telescope (CDK20). All observations were carried out in the optical band. The observation log is presented in Table~\ref{tab:T1_v1405cas}.
\subsubsection{The 2m Himalayan Chandra Telescope (HCT)}
In this study, twelve of the twenty-seven spectra are of medium resolution (\(R \sim 1400\)–\(2200\)), obtained with HFOSC on the HCT using Grism~7 (3800–7000~\AA) and Grism~8 (5200–9000~\AA), with a slit width of 1.92~arcsec (167~\(\mu\)m) and a slit height of about 11~arcmin.  All observations were conducted using a 2k~\(\times\)~4k CCD. Lamp spectra (FeAr and FeNe) were obtained immediately after each science frame for wavelength calibration. On the same night, the spectrophotometric standard stars Feige~34 (O-type subdwarf) and Feige~110 (OB-type subdwarf) were observed through both grisms \citep{1990ApJ...358..344M, 2002ApJS..140...37F}. Details of the observations are provided in Table~\ref{tab:T1_v1405cas}.

The optical spectra were reduced using a standard pipeline based on the Image Reduction and Analysis Facility (\textsc{iraf}) package\footnote{\url{http://iraf.noao.edu/}} \citep{1993ASPC...52..173T}. First, the master bias frame was subtracted from each science frame, and cosmic rays were removed from the bias-subtracted frames. One-dimensional spectral data were then extracted from the 2D images using the \textsc{apall} task in \textsc{iraf}. Wavelength calibration was performed using dispersion-corrected lamp spectra (FeAr for Grism 7 and FeNe for Grism 8). Flux calibration was carried out using the atmospheric extinction function and detector sensitivity function derived from standard star data.  

To account for interstellar extinction, all spectra were dereddened using a reddening value of \(E(B-V) = 0.53\) \citep{2022ATel15796Munari,2023arXiv230204656VValisa}. All spectra, including those shown in  Fig.~\ref{fig:specg7}, \ref{fig:graph9}, \ref{fig:linepnew}, and \ref{fig:graph13}, were flux-calibrated using spectrophotometric standard stars observed nightly. For Figures \ref{fig:lr} and \ref{fig:cloudy_best_fitting_model}, we further normalized the spectra to H$\beta$ to enable direct comparison of relative line strengths and profiles across epochs.

\subsubsection{C11 SCT, C14 SCT and CDK20}
Fifteen of the twenty-seven medium- and low-resolution spectra analyzed in this work were obtained from the Astronomical Ring for Access to Spectroscopy (ARAS) database\footnote{\url{https://aras-database.github.io/database/novae.html}} \citep{2019Teyssier, 2024CoSka..54b.107T}. These low-resolution spectra were obtained using a spectrograph of a slit width 23~\(\mu\mathrm{m}\). The ARAS symbiotic project serves as an archival repository and comprises a network of small telescopes, ranging from 20 to 60 cm in diameter, equipped with spectrographs that provide resolutions spanning from R $\sim$ 500 to 15,000, covering wavelengths from 3600 to $\sim$9000~\AA. These spectra were acquired with the C11, C14, and CDK20 instruments, located in the UK, Slovakia, and USA. The C11--UK spectrum (covering $\sim$3900--7380~\AA) was obtained at a spectral resolution of 1030, the C11 and C14 Slovakia spectra (covering $\sim$3800--7550~\AA) at a resolution of 976, and the CDK20--USA spectrum (covering $\sim$3730--7300~\AA) at a resolution of 1059.  The observational details and corresponding observatories are outlined in Table \ref{tab:T1_v1405cas}. The spectra were dark subtracted and flat divided, and reduced using the Integrated Spectroscopic Innovative Software (\textsc{ISIS})\footnote{\url{http://www.astrosurf.com/buil/isis-software.html}}, written by C. Buil \citep{2019Teyssier}.  

\begin{deluxetable}{cccccccc}  
	\label{tab:T1_v1405cas} 
	\tablecaption{Log of optical spectral observation of V1405 Cas.}
	\tablehead{
		\colhead{MJD} & \colhead{t$^a$(days)} & \colhead{Observer} & \colhead{Telescope} & \colhead{Spectrograph/Camera} & \colhead{R$^{b}$}& \colhead{Coverage (\AA)} & \colhead{TTE$^{c}$(s)}
	} 
	\startdata
	59293.342&1.918&DBO$^{d}$&C11 SCT$^{j}$&LISA / SXVR-H694&1030&3900-7380&2718\\
	59315.092&23.576&PAD$^{e}$&C11 SCT$^{k}$&LISA / Atik 460EX&856&3800-7590&1846\\ 
	59383.457&92.003&PAD$^{e}$&C11 SCT$^{k}$&LISA / Atik 460EX&970&3800-7550&738\\ 
	59461.353&169.929&PAD$^{e}$&C14 SCT$^{l}$&LISA / Atik 460EX&1076&3900-7380&1519\\
	59531.391&239.967&PAD$^{e}$&C14 SCT$^{l}$&LISA / Atik 460EX&976&3800-7550&1231 \\
	59575.224&283.800&PAD$^{e}$&C14 SCT$^{l}$&LISA / Atik 460EX&911&3850-7550&4825\\
	59595.289&303.865&PAD$^{e}$&C14 SCT$^{l}$&LISA / Atik 460EX&888&3900-7550&3015\\ 
	59630.309&338.885&PAD$^{e}$&C14 SCT$^{l}$&LISA / Atik-160ex&985&3850-7500&3015\\ 
	59662.608&371.174&PAD$^{e}$&C14 SCT$^{l}$&LISA / Atik-460EX&942&3850-7400&3012\\ 
	59709.474&418.050&PAD$^{e}$&C14 SCT$^{l}$&LISA / Atil460ex&1051&3900-7400&3131\\ 
	59723.510&432.086&PAD$^{e}$&C14 SCT$^{l}$&LISA / Atik 460EX&963&3900-7400&1205\\
	59745.500&454.076&PAD$^{e}$&C14 SCT$^{l}$&LISA / Atil460ex&943&3900-7400&3014\\
	59771.456&480.032&PAD$^{e}$&C14 SCT$^{l}$&LISA / Atil460ex&1035&3950-7400&1203\\
	59786.505&495.081&YAM$^{f}$&C14 SCT$^{l}$&LISA / Atil460ex&1129&3900-7400&3013\\
	59814.658&523.234&GRH$^{g}$&HCT$^{m}$&HFOSC - Gr7&1300&4000-7000&60\\
	59814.670&523.246&GRH$^{g}$&HCT$^{m}$&HFOSC - Gr8&2200&5200-9000&60\\
	59892.609&601.185&FAS$^{h}$&CDK20$^{n}$&LISA / Atik414ex&1059&3730-7300&2889\\ 
	59916.660&625.236&GRH$^{g}$&HCT$^{m}$&HFOSC - Gr7&1300&4000-7000&600\\
	59916.675&625.251&GRH$^{g}$&HCT$^{m}$&HFOSC - Gr8&2200&5200-9000&600\\
	59974.577&683.153&SK$^{i}$&HCT$^{m}$&HFOSC - Gr7&1300&5200-9000&1200\\
	59974.603&683.179&SK$^{i}$&HCT$^{m}$&HFOSC - Gr8&2200&5200-9000&600\\
	60165.857&874.433&GRH$^{g}$&HCT$^{m}$&HFOSC - Gr7&1300&4000-7000&1500\\
	60165.891&874.467&GRH$^{g}$&HCT$^{m}$&HFOSC - Gr8&2200&5200-9000&1200\\
	60271.763&980.390&GRH$^{g}$&HCT$^{m}$&HFOSC - Gr7&1300&4000-7000&1200\\
	60271.779&980.355&GRH$^{g}$&HCT$^{m}$&HFOSC - Gr8&2200&5200-9000&1200\\
	60342.559&1051.135&GRH$^{g}$&HCT$^{m}$&HFOSC - Gr7&1300&4000-7000&1200\\
	60342.602&1051.178&GRH$^{g}$&HCT$^{m}$&HFOSC - Gr8&2200&5200-9000&1200\\
	\enddata
	\tablecomments{$^{(a)}$Number of days counted from $t_0$ (2021 March 19.424 UT, JD 2459291.924),  $^{(b)}$Resolution, $^{(c)}$Total Time of Exposure,  $^{(d)}$David Boyd, $^{(e)}$Pavol A. Dubovsky, $^{(f)}$Yamamoto, $^{(g)}$Gesesew R. Habtie, $^{(h)}$Forrest Sims, $^{(i)}$Subhajit Kar, $^{(j)}$\textit{Celestron 11-inch Schmidt--Cassegrain Telescope} in England, $^{(k)}$\textit{Celestron 11-inch Schmidt--Cassegrain Telescope} in Slovakia, $^{(l)}$\textit{Celestron 14-inch Schmidt--Cassegrain Telescope} in Slovakia,,  $^{(m)}$\textit{Himalayan Chandra Telescope} in India, $^{(n)}$\textit{PlaneWave 20-inch Corrected Dall--Kirkham Telescope} in USA. }
\end{deluxetable}

\section{Results and Discussion}\label{sec:results}
\subsection{Optical light curves}\label{lc}
\begin{figure}[htbp!] 
	\centering    
	\includegraphics[width=0.55\textwidth]{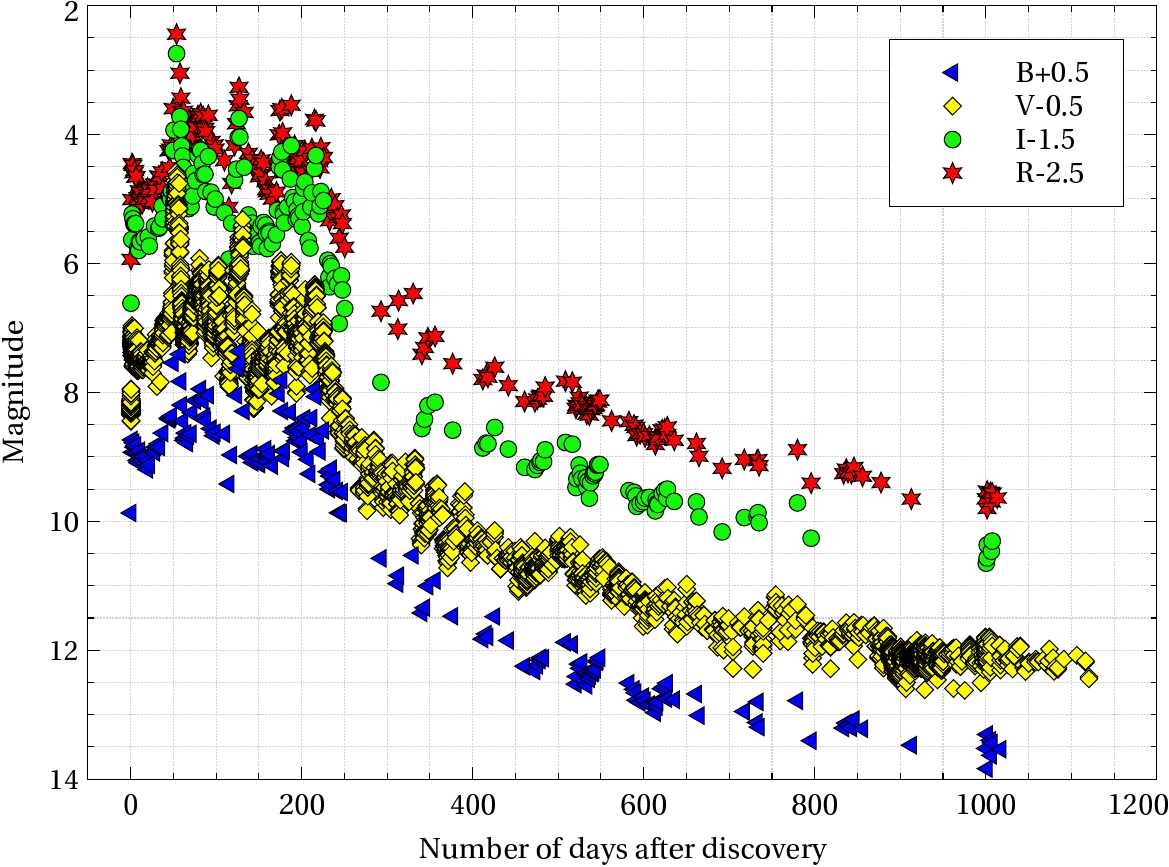} 
	\caption{\textit{BVRI} Light curves of Nova Cas 2021 generated using optical data obtained from AAVSO. Offsets are applied to the R, B, and I bands for the sake of clarity.}
	\label{fig:lightcurvenher21}
\end{figure}

The optical \textit{BVRI} light curve of nova V1405 Cas over the initial 1150 days, constructed using the AAVSO data, is shown in Fig.~\ref{fig:lightcurvenher21}. The peak brightness, $V = 5.08 \pm 0.06$, was reached on 2021 May 10.297 UT (JD 2459344.797). Therefore, we designate this date as the reference time for the primary maximum. Apart from the primary maximum, we observed multiple Maxima of the nova at different times, but all of them were less bright than the primary maximum. Due to the presence of these multiple rebrightenings, the $t_2$ and $t_3$ decline times are determined from the final decline phase following the last maximum, which occurs at $\mathrm{JD}\,2459507.243681$, after which the nova exhibits a monotonic fading. Using this approach, we derive $t_2 \sim 165$ days and $t_3 \sim 175$ days. The obtained $t_2$ value is in good agreement with \citep{2024MNRASAydi, 2025NatAs.tmp..249A}.   
 
Similar to the nova V1405~Cas, several other CNe have exhibited rebrightenings prior to their final decline in brightness \citep{1989AJ.....98..297W, 1998A&A...338.1006I, 2011PASJ...63..159T}. Such novae are classified as J-type systems in the light-curve scheme of \cite{2010AJ....140...34S}. Various explanations have been proposed to account for these rebrightening events. \citet{2011PASJ...63..159T} suggested that the rebrightening is primarily due to an enhancement of the continuum emission rather than an increase in emission-line flux, likely associated with the re-expansion of the photosphere following the initial eruption. Alternatively, \citet{2009ApJ...701L.119P} proposed that the rebrightenings are driven by instabilities in hydrogen shell burning. Most novae that display such rebrightenings are categorized as slow or moderately slow novae \citep{2011PASJ...63..159T}.   

According to the Gaia parallax-based measurement, the distance to nova V1405~Cas has been estimated to be \(1.73^{+0.08}_{-0.07}\)~kpc \citep{2023ApJTaguchi}. We adopted this value for further analysis in this paper, as Gaia parallax typically provides the most direct and reliable distance estimates. 

\subsubsection{White Dwarf and Secondary Star Masses} \label{wdssm}
Using the $t_3$ value in the $V$ band estimated in Section~\ref{lc}, and applying the $t_3$---$M_{\mathrm{WD}}$ relation from \citet{1992ApJLivio}, we estimated the mass of the WD in nova V1405~Cas. Using these results, and adopting the same calibration constant derived for V1500~Cyg ($C = 51.3$), we obtain a WD mass of $\sim 0.7\~M_{\odot}$. This value is consistent with the estimate of \citet{2023ApJTaguchi}, who suggested a WD mass potentially below $1.1\~M_{\odot}$. In most cases, such low masses are typically associated with CO WDs.

Based on post-outburst optical light curves obtained by \textit{TESS}, \citet{2025AAS...24521005B, 2021RNAAS...5..150S} confirmed that the orbital period of nova V1405~Cas is $\sim$ 4.52 hours (or $\sim$0.19 days), with a WD spin period of about 1.23 hours. This relatively short orbital period offers important insights into the properties of the secondary (donor) star, including its mass, radius, effective temperature, luminosity, and spectral type. Using the semi-empirical donor sequence established by \citet{ warner1995, 2011ApJS..194...28K}, we estimate the secondary’s mass and radius to be $M_{\rm sec} \approx 0.43\~M_{\odot}$ and $R_{\rm sec} \approx 0.48\~R_{\odot}$, respectively. Furthermore, using the estimated parameters and applying Kepler’s third law, we calculate the binary separation as
	\begin{equation}
		a~=~3.53\times10^{10}M_{WD}^\frac{1}{3}\left(1+\frac{M_{\rm sec}}{M_{WD}}\right)^{1/3}P_{\rm orb}^{2/3}~\text{cm},
	\end{equation}  
	yielding $a \approx 1.01 \times 10^{11}~\text{cm}~(\approx 1.45~R_{\odot})$. 

\subsection{Spectral Evolution }\label{sec3}
\begin{figure}[h!]
	\centering
	\includegraphics[scale=0.6]{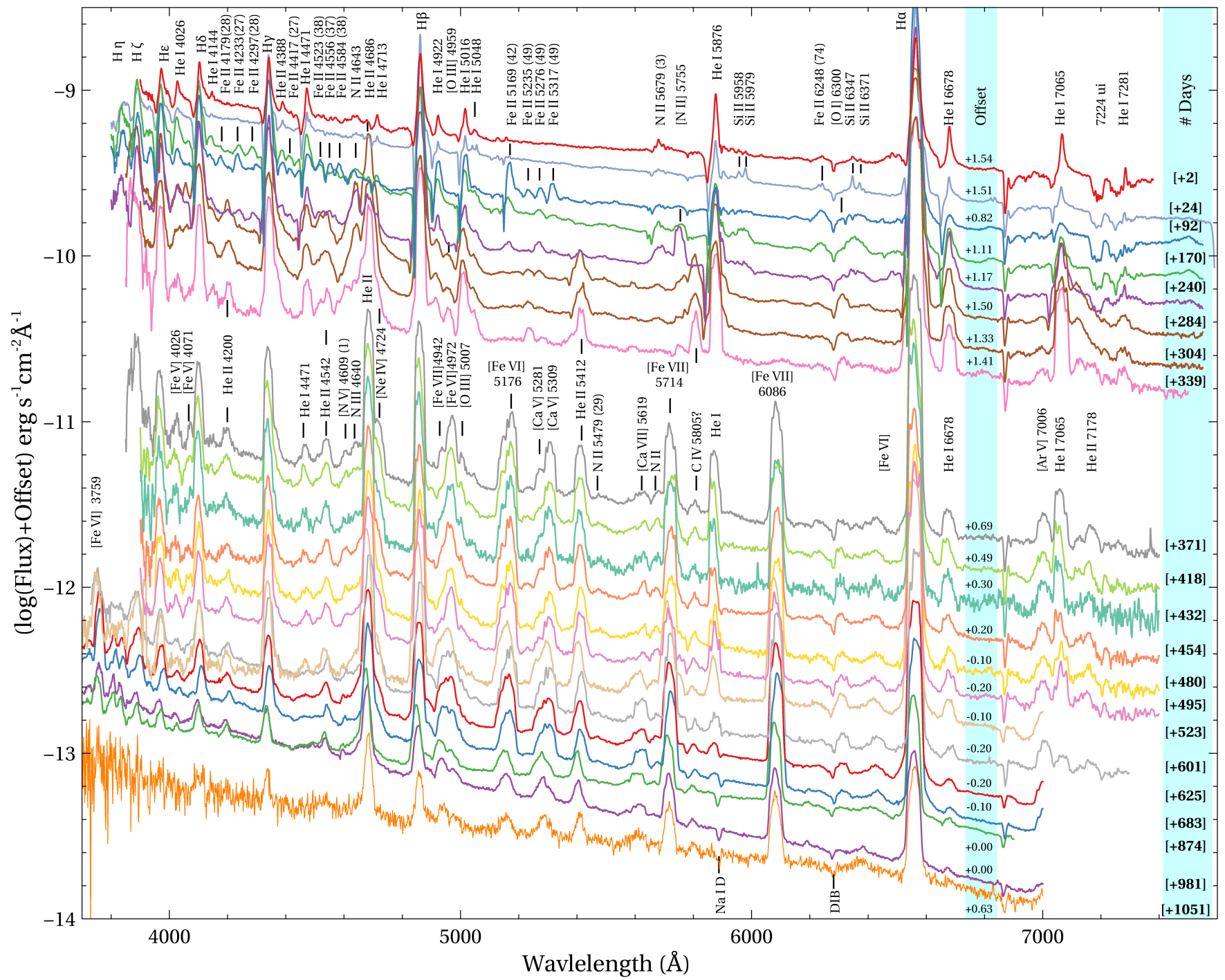}
	\caption{	Reddening-corrected and flux-calibrated spectra of V1405~Cas obtained at twenty-one epochs, with the corresponding days indicated on the right. For clarity, most spectra have been vertically offset as shown in the figure.  The vertical axis represents the flux on a logarithmic scale, while the horizontal axis denotes the wavelength in \AA. Notably, prominent helium and iron lines appeared in the spectra around day~92.}	
	\label{fig:specg7}
\end{figure}

\begin{figure}
	\centering
	\hspace*{0.2cm}	\includegraphics[scale=0.5]{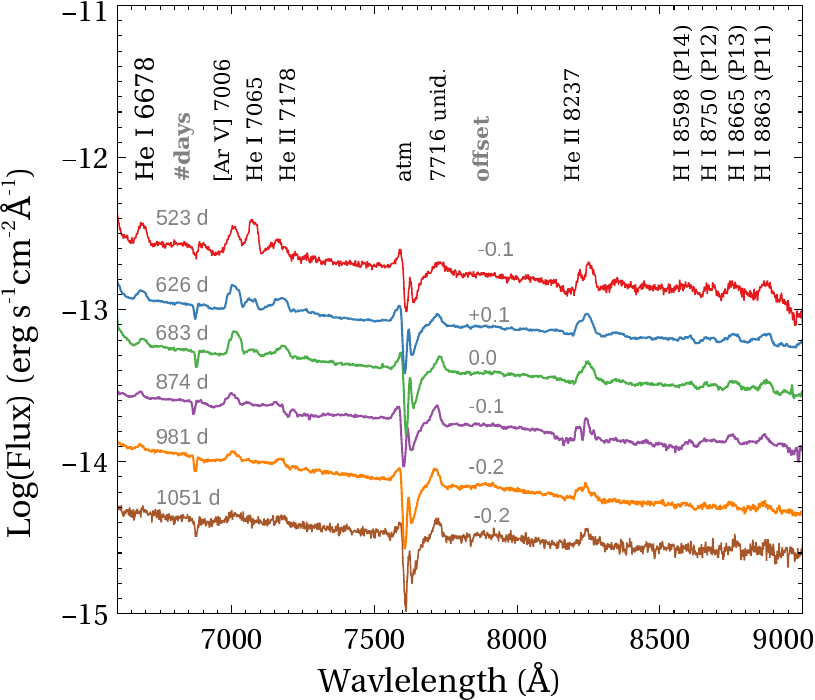}
	\caption{Gr-8 spectra of V1405 Cas span from 6600 to 9000 \AA. The dates following the outburst are indicated in grey on the left side within the box. Each emission line is offset and labeled in grey at the center. The vertical scale is logarithmic to enhance the visibility of fainter features. The term 'atm' denotes atmospheric absorption observed around 6876 \AA. }
	\label{fig:graph9}
\end{figure}

The evolution of the optical spectra of V1405~Cas from day 2 up to 1051 days after the outburst is shown in Figures~\ref{fig:specg7} and \ref{fig:graph9}. Fig.~\ref{fig:specg7} displays a set of 21 spectra obtained from various telescopes, while Fig.~\ref{fig:graph9} presents the remaining six spectra acquired using the Grism~8 setup of HCT (see Table~\ref{tab:T1_v1405cas}). The Gr-8 spectra cover only the 6600--9000~\AA\ range, since the bluer region is already covered by Gr-7.  This region of the spectra is dominated by various strong telluric absorption lines and some other weak emission lines including Balmer, helium and oxygen lines.

On days +2, 24, 92, 170, and +284, the spectra are dominated by strong hydrogen and helium emission lines with superimposed P~Cygni absorption features (Figs.~\ref{fig:specg7} and \ref{fig:linepnew}). During these epochs, the He~\textsc{I} lines at 4028, 4471, 4922, 5016, 5876, 6678, and 7065~\AA\ are particularly prominent, appearing both strong and broad. As the nova evolves, these lines gradually weaken, consistent with the increasing dominance of higher ionization stages of helium. In addition to the recombination lines discussed above, several Fe~\textsc{II} multiplets and Si~\textsc{II} doublets emerged in the spectra, including Fe~\textsc{II} multiplets 27, 29, 37, 38, 41, 42, 46, 49, 55, 73, and 74, as well as the Si~\textsc{II} doublet at 6347 and 6371~\AA. \citet{2023arXiv230204656VValisa} reported that these Fe~\textsc{II} multiplet lines formed relatively early, from day +31.4 onward, coinciding with the weakening of the He~\textsc{I} lines. In our dataset (see Fig.~\ref{fig:specg7}), the iron lines became clearly visible by day +92, although some features---such as Fe~\textsc{II}~6248~\AA---appear to be present as early as day +24 or even earlier. Similarly, the Si~\textsc{II} doublets were clearly detected in the early spectra (days +2 and +24) of Nova~V1405~Cas (see Fig.~\ref{fig:specg7}). Unlike the broad Balmer and Fe~\textsc{II} features tracing the fast-expanding ejecta, the Si~\textsc{II} lines appeared relatively narrow, suggesting formation in a denser and slower-moving region of the system, possibly associated with clumpy ejecta or circumbinary material. A similar situation has been observed in Nova~LMC~2004 \citep{2014A&A...569A..84M}. These low-ionization transitions imply that the gas remained comparatively cool and dense, favoring the survival of singly ionized silicon \citep{1988MNRAS.232..507R}. The presence of narrow Si~\textsc{II} emission further supports a structured ejecta configuration in which high-velocity outflows coexist with slower, confined components \citep{2014A&A...569A..84M, 2016ApJ...830...30A, 2020ApJ...905...62A, 2025NatAs.tmp..249A}.

On days +24, and +92, the P~Cygni profiles of H$\alpha$, H$\beta$, and He~\textsc{I}~5876~\AA\ exhibit a slight redshift relative to their earlier velocities.   The trend then reversed around day~+170, when the P~Cygni profiles began to blueshift substantially, continuing to accelerate until the absorption components became unresolved at the spectral resolution of our observations by day +339. This evolution reflects the progressive acceleration of the ejecta along the line of sight and the development of a fast-moving outflow. These strong P~Cygni profiles persisted at least up to day~+304 after the outburst, marking one of the longest durations for such absorption features observed in novae. Their persistent presence suggests that the nova wind or outflow remained active for an extended period, possibly sustained by continued energy injection from residual hydrogen burning on the WD surface or by delayed mass-ejection episodes, both of which are commonly observed in very slow novae \citep{2020ApJ...905...62A, 2025NatAs.tmp..249A}. In contrast, such features typically disappear much earlier in faster novae, such as RS~Oph~2021 and V1674~Her \citep{2022MNRASpandey, 2024MNRASHabtie}. Moreover, we note that the weakening of the P~Cygni absorption is accompanied by a broadening of the emission components (increased FWHM) and a decline in their integrated flux. This behavior is consistent with the expanding ejecta becoming progressively more diffuse and transitioning toward a lower-density, optically thin regime \citep{1991ApJ...376..721W, 1993AJ....106.2408S, 2008Bodebook}. 

One of the most striking spectral changes, which is also common in many other novae, is the sudden and dramatic rise in the intensity of the He\,\textsc{II}\,4686\,\AA\ emission line \citep{2022ATel15540Woodward, 2023ATel16089....1W, 2024ATel16496_Habtie, 2025AAS...24521005B}. The line intensity evolved substantially, eventually surpassing that of H$\beta$, consistent with the behaviour reported by \citet{1992AJ....104..725W}, who noted similar enhancements in nova spectra during later evolutionary phases. The He\,\textsc{II}\,4686\,\AA\ line first became slightly detectable around day~170, with a line flux of \(3.28\times10^{-11}\)~erg\,cm\(^{-2}\)\,s\(^{-1}\) and an FWHM of 932~\kms. Approximately 100 days later, on day~284, its flux had abruptly increased to \(5.59\times10^{-10}\)~erg\,cm\(^{-2}\)\,s\(^{-1}\), accompanied by a marked broadening to an FWHM of 2626~\kms (see Table~\ref{tab:FWHM01_v1405}).  Furthermore, higher excitation He \textsc{ii} lines---specifically 4542, 4200, and 5412~\AA---started to appear in the subsequent spectra. Contrary to the trend of such increasing high-ionization features, lower-ionization species like He \textsc{i} showed a marked decline. Similarly, as the nova evolved, the Fe II lines gradually weakened and were replaced by a range of higher-ionization coronal features. By day~+371, prominent coronal lines, including [Fe~\textsc{VI}]~5147, 5179, and 5277~\AA, [Fe~\textsc{VII}]~4942, 5721, and 6086~\AA, [Ca~\textsc{V}]~5281~\AA, and [Ca~\textsc{VII}]~5619~\AA, had emerged, indicating that the nova had already entered the coronal phase by this epoch \citep{2022ATel15540Woodward,2023ATel16089....1W,2023ATel15934....1R}. However, by this time, some classical nebular lines, such as [O~\textsc{III}]~4959~\AA, [Ne~\textsc{III}]~3869 and 3968~\AA, and [Ne~\textsc{IV}]~4724~\AA, remained weak or absent, suggesting that the ejecta still had relatively high densities and/or that the nebular emission was not yet fully developed.\citet{2023AAS...24127403P} reported that this behavior persisted until at least day~+531, while \citet{2024BlgAJ..40...13M} found that the nebular phase became more apparent by day~+547.

\subsubsection{Line Profile}\label{lp} 
\begin{figure}[h!]
	\centering
	\includegraphics[scale=0.45]{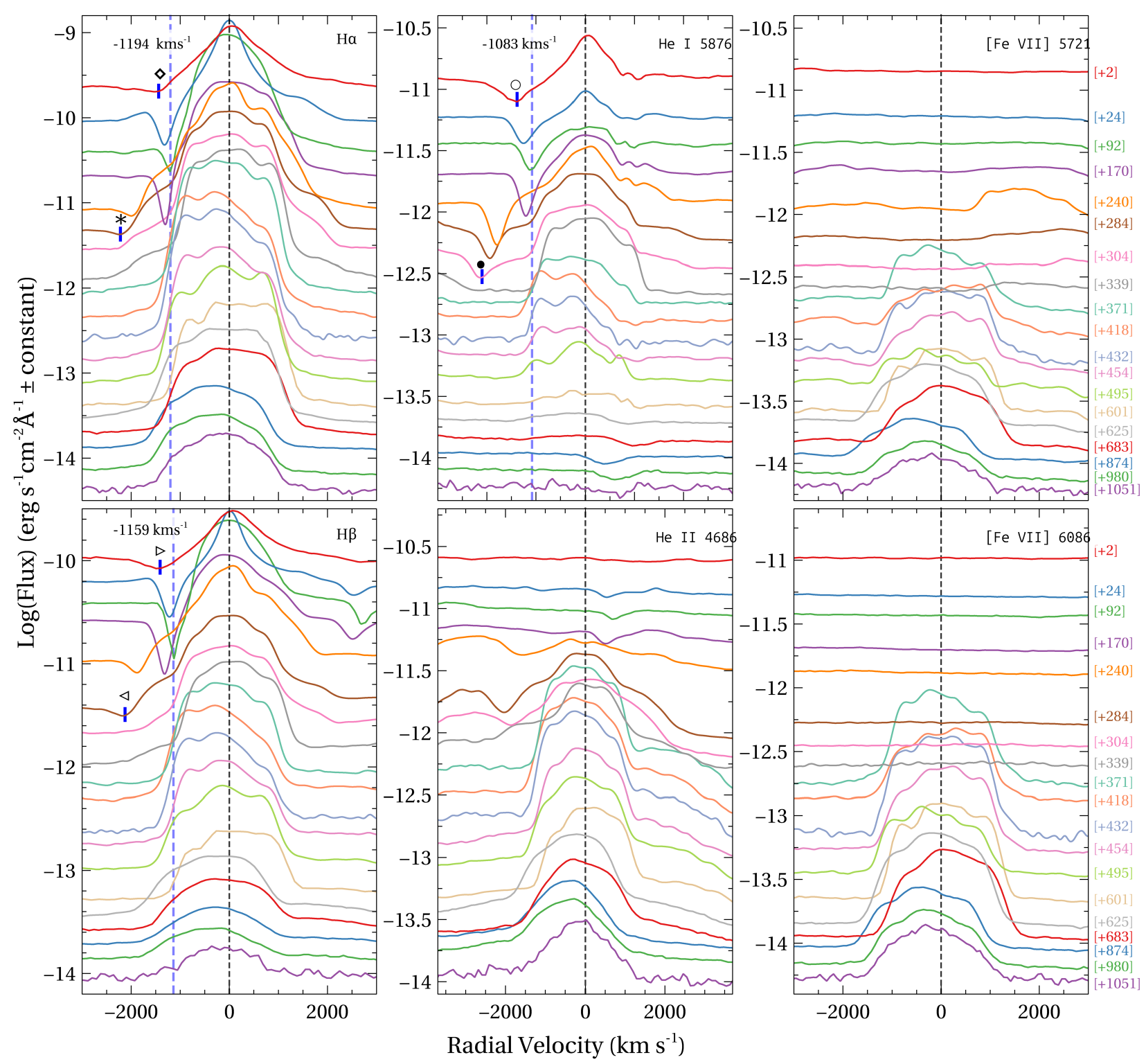} 
	\caption{ Line profiles of H$\alpha$, H$\beta$, H$\gamma$, He \textsc{II} 4686 \AA, [Fe \textsc{VII}] 5721 \AA, and [Fe \textsc{VII}] 6086 \AA. In the H$\alpha$ panel, the symbols "$\ast$" and "$\diamond$" mark the central velocities of the P Cygni absorption minima at $\sim$ $-1441$ km s$^{-1}$ (day +2) and $-2221$ km s$^{-1}$ (day +284), respectively. In the H$\beta$ panel, the symbols "$\triangleright$" and "$\triangleleft$" denote velocities of $-1437$ km s$^{-1}$ and $-2153$ km s$^{-1}$, respectively. Similarly, in the He \textsc{I} 5876 \AA~panel, the symbols "$\circ$" and "$\bullet$" correspond to velocities of $-1389$ km s$^{-1}$ and $-2123$ km s$^{-1}$, respectively. However, the central velocities are subject to uncertainties of about 10--20\%, depending on their strength and degree of blending. The blue dashed lines indicate the minimum P Cygni absorption velocities in the H$\alpha$, H$\beta$, and He \textsc{I} 5876 \AA~profiles, while the black dashed lines in all panels mark the heliocentric rest velocity. The numbers on the right-hand side of each panel indicate the number of days since discovery. The left ordinate and the bottom abscissa represent the logarithmic line flux (\ergscmA) and the heliocentric velocity (km s$^{-1}$), respectively.} 
	\label{fig:linepnew}
\end{figure}

Fig. \ref{fig:linepnew} presents the line profiles of prominent emission lines of nineteen different epochs. For each line profile heliocentric correction has been applied for the purpose of accounting the Earth's motion relative to the Sun. The observed full width at half maximum (FWHM$_{\text{obs}}$) values of the emission lines were corrected for the instrumental resolution to derive their intrinsic widths (FWHM$_{\text{true}}$), using $\text{FWHM}_{\text{true}} = \sqrt{\text{FWHM}_{\text{obs}}^{2} - \text{FWHM}_{\text{inst}}^{2}}$. The instrumental broadening (FWHM$_{\text{inst}}$) was estimated from the spectral resolution. This correction ensures that the derived line widths represent the physical velocity dispersion of the emitting gas rather than being dominated by instrumental effects.  The measured values of  flux, FWHM, and equivalent width (EW) of the most prominent emission lines are presented in Table~\ref{tab:FWHM01_v1405}.  

\begin{figure*}
	\centering
		\includegraphics[scale=0.35]{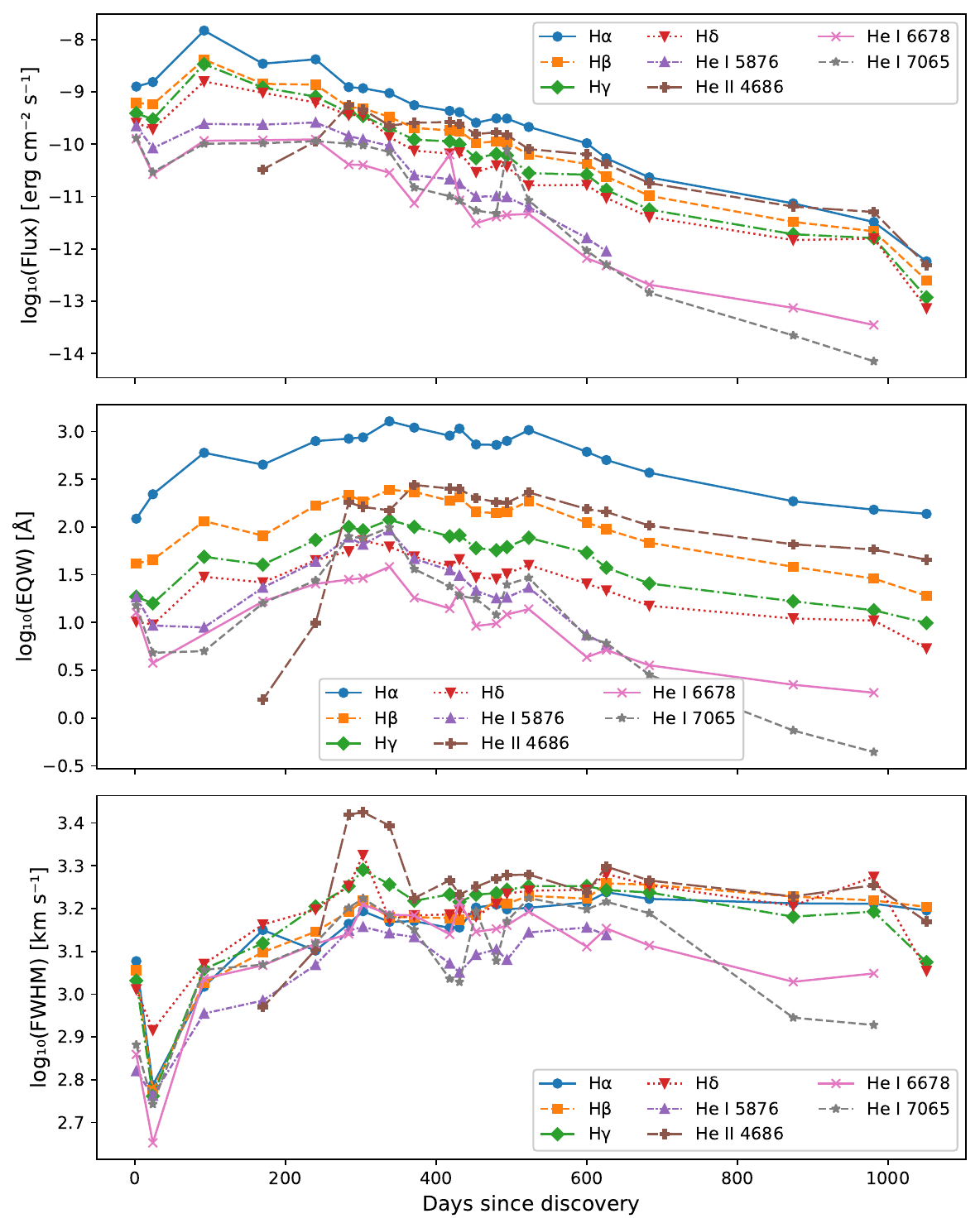}\\  
	\caption{ Temporal evolution of the most prominent emission lines—H$\alpha$, H$\beta$, H$\gamma$, H$\sigma$, He\,\textsc{I}\,5876, 6678, 7065~\AA{}, and He\,\textsc{II}\,4686 ~\AA{}.  The upper panel shows the flux evolution of the aforementioned lines over the first 1050~days. Most of the line fluxes gradually increased until day~92, followed by a steady decline thereafter.  The middle panel presents the evolution of the EW with time, where the maximum value was reached by H$\alpha$ around day~339.  The bottom panel illustrates the variation of the FWHM with time.To the contrary of the flux and EW, the FWHM values initially decreased around day~24, followed by a gradual increase. The most prominent broadening was observed in the He\,\textsc{II}\,4686~\AA{} line. }	
	\label{fig:graph13}
\end{figure*}

During the first $\sim$200 days after the outburst of V1405~Cas, the optical light curve exhibited a series of multiple rebrightening events (see Sec.~\ref{lc}). These episodes were accompanied by a noticeable rise in the flux of the prominent Balmer lines up to day~171, followed by a gradual decline thereafter (see Fig.~\ref{fig:graph13}). This behaviour suggests that each rebrightening phase was associated with renewed energy injection into the system. Such episodes are commonly linked to instabilities in the hydrogen-burning shell on the WD surface, late-time mass ejection, or delayed wind-driven outflows \citep{2009arXiv0904.2228K,2016ApJHachisu}. These mechanisms can enhance the ionizing radiation from the central WD, increasing the Balmer line fluxes through more efficient recombination in the ejecta \citep{2004A&ACassatella}. Similarly, the EWs of the Balmer lines continued to increase up to day~338. As shown in Table~\ref{tab:FWHM01_v1405}, the integrated flux of all emission lines steadily decreases with time, while their EWs increase. Since the EW measures the line flux relative to the continuum, this behavior indicates that the continuum flux declines more rapidly than the line emission. Thus, the observed increase in EW reflects the faster fading of the continuum rather than intrinsic line strengthening. In contrast, the FWHM values of all lines exhibited a pronounced decrease on day~24, followed by a gradual increase up to day~303 (see Fig.~\ref{fig:graph13}). 

The He\,\textsc{I} lines, including 5876, 6678, and 7065~\AA, gradually weakened over time, with He\,\textsc{I}\,5876\,\AA\ becoming nearly undetectable by day +683 (see Fig.~\ref{fig:graph13}). While He~I~5876 and 7065~\AA\ are triplet transitions and show broadly similar behavior, the singlet transition He~I~6678~\AA\ evolves somewhat differently, likely due to its distinct excitation mechanism and lower optical depth sensitivity. The triplet lines are strongly influenced by the metastable $2^{3}$S level, which can retain a significant population of helium atoms and enhance collisional excitation and radiative trapping, making these transitions particularly sensitive to density and optical depth effects in the ejecta \citep{1999ApJ...514..307B, 2006agna.book.....O}. In contrast, the singlet He~I~6678~\AA\ line is less affected by the metastable-state population and therefore responds differently to the evolving physical conditions as the ejecta expand and become optically thinner. As seen in Fig.~\ref{fig:graph13}, the relative strengths and profiles of the singlet and triplet lines diverge with time, suggesting a progressive decrease in optical depth during the late decline phase. Other prominent lines, such as H$\alpha$, H$\beta$, H$\gamma$, and He~II~4686~\AA, exhibit a more gradual decline, consistent with the typical spectral evolution of novae.

\begin{figure}[h!]
	\centering
	\includegraphics[scale=0.6]{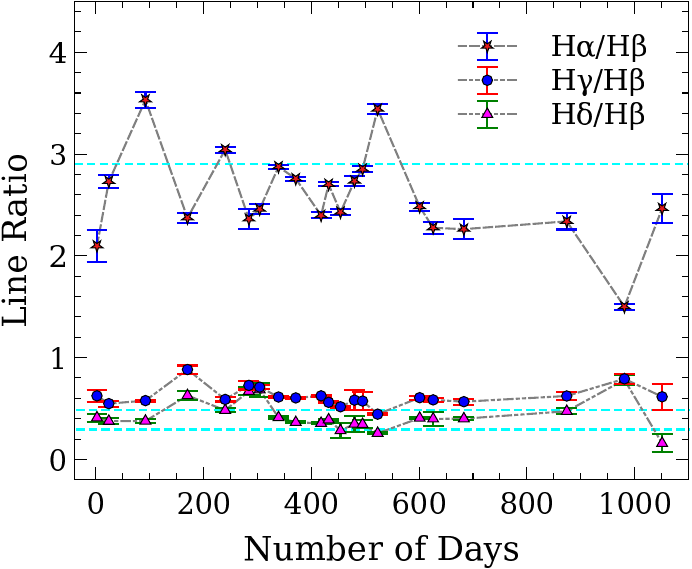}
	\caption{ Evolution of the Balmer line ratios relative to H$\beta$. The horizontal lines represent the expected Case\,B values at an electron temperature of $10{,}000~\mathrm{K}$: H$\alpha$/H$\beta = 2.85$, H$\gamma$/H$\beta = 0.47$, and H$\delta$/H$\beta = 0.26$, listed from top to bottom \citep{1987MNRAS.224..801H,1989agnaOsterbrock}.}	
	\label{fig:lr}
\end{figure}

Fig.~\ref{fig:lr} presents the temporal evolution of the flux ratios H$\alpha$/H$\beta$, H$\gamma$/H$\beta$, and H$\delta$/H$\beta$ from day~+2 to~+1051. The three dashed horizontal reference lines, from top to bottom, represent the theoretical Balmer line ratios expected for Case~B recombination at $T_{\mathrm{e}} = 10{,}000$~K, as calculated by \citet{1987MNRAS.224..801H,1989agnaOsterbrock}. The observed line ratios show the closest agreement with the theoretical Case~B values between approximately day~200---when the nova entered its gradual decline in brightness---and day~523, corresponding to the nebular phase and the early stage of the coronal phase. During this interval, the ejecta are sufficiently optically thin, making the Case~B recombination assumption appropriate for typical nebular densities and temperatures. Although the deviations from the theoretical Case~B Balmer ratios are not large, most of the measured H$\alpha$/H$\beta$ values lie slightly below the predicted value, whereas the observed H$\gamma$/H$\beta$ and H$\delta$/H$\beta$ ratios tend to fall marginally above their corresponding Case~B expectations. Such a pattern can arise when the lower-order Balmer transitions (particularly H$\alpha$) experience mild optical-depth or self-absorption effects, which preferentially suppress H$\alpha$ relative to H$\beta$, while the higher-order Balmer lines remain comparatively optically thin and therefore appear slightly enhanced \citep[e.g.,][]{1975MNRAS.171..395N,1980ApJS...42..351D}. Similar behaviour has been reported in nebular and nova ejecta under conditions where Balmer lines deviate modestly from pure Case~B recombination \citep[e.g.,][]{1989agnaOsterbrock}. 

The uncertainties in the emission-line ratios were estimated using standard statistical methods based on repeated measurements. The resulting line-flux uncertainties were then propagated into the derived ratios using standard Gaussian error propagation, under the assumption that the uncertainties in different emission lines are independent and normally distributed. The H$\alpha$/H$\beta$, H$\gamma$/H$\beta$, and H$\delta$/H$\beta$ ratios were found to have typical uncertainties ranging from 1\% to 16\%. 

\subsection{Photo-ionization Model}\label{pma}
We use the photoionization code \textsc{CLOUDY}\footnote{\url{https://trac.nublado.org/}} (v23.01; \citealt{2023RNAAS...7..246G}), to model the emission line spectra of the nova V1405 Cas. Further details about our use of this code are provided in our previous publications \citep{2022MNRASpandey, 2024MNRASHabtie, 2024MNRAS.529..917H, 2024arXivHabtie, 2025MNRAS.537.2046H, 2024asi..confO..51H}, which we recommend for additional information.

We made the assumption that the surface of the central WD emits ionizing blackbody radiation with a temperature $T_{BB}$ and a bolometric luminosity $L_{bol}$, which irradiates a spherical gas geometry expanding at a velocity of $v_{exp}$. The spatial dimensions of the gas are defined by an inner radius $\mathrm{r_{in}}$ and an outer radius $\mathrm{r_{out}}$. For synthetic model spectra, we incorporate these input parameters along with the abundances of elements observed in emission lines, while other elements are maintained at their solar values as described by \citet{2010ApGrevesse}. 

The density of the ejecta is set by the total hydrogen number density, n(H) [$cm^{-3}$ ] given by,
\begin{equation}
	\mathrm{n_H} = \mathrm{n_{H^0}} + \mathrm{n_{H^+}} + \mathrm{2n_{H_2}} + \mathrm{\sum_{other}n_{H_{other}}} \mathrm{cm^{-3}},
\end{equation}
where, $\mathrm{n_{H^0}}$, $\mathrm{n_{H^+}}$, $\mathrm{2n_{H_2}}$, and $\mathrm{n_{H_{other}}}$ respectively represent hydrogen in its neutral, ionized, molecular, and all other hydrogen-bearing molecular forms. The elemental abundances of the ejecta, relative to hydrogen, are determined by the abundance parameter. Additionally, we employ a radius-dependent density distribution governed by a power-law function, $\rho \propto r^\alpha$. The level of clumpiness, indicating the gas proportion within the total volume, is determined by the filling factor parameter also denoted as $\textrm{f(r)}$. \textsc{cloudy} employs the following two distribution functions to express the variations in density and filling factor within the ejecta: 

\begin{equation}\label{ndff}
\mathrm{n(r)=n(r_{in})x^{\alpha}}	\qquad \text{and} \qquad \mathrm{f(r)=f(r_{in})x^{\beta}}, 
\end{equation}
where $\mathrm{\alpha}$ and $\mathrm{\beta}$ represent the power-law indices and $\mathrm{x}$ represents the ratio of $\mathrm{r_{in}}$ to $\mathrm{r_{out}}$ of the shell. For this study we used; $\mathrm{\alpha}=-2$ and $\mathrm{\beta}=0$.  The value of $\mathrm{\alpha}=-2$ was chosen to maintain a constant mass per unit volume throughout the model shell ($\dot{\textrm{M}}=$ const.) and a linear velocity law ($\textrm{v} \propto \textrm{r}$). On the other hand, we chose $\mathrm{\beta}=0$ because the filling factor of novae ejecta typically reaches a maximum of 0.1 during the outburst phase and gradually decreases as the ejecta disperses over extended distances from the source \citep{2006A&AEderoclite, shore_2008, 2022ApJ...925..187P}.

The inner and outer shell radii were estimated using the minimum and maximum expansion velocities derived from the \textsc{FWHM} of the emission lines together with the elapsed time since the outburst. Directly equating the expansion velocity with the \textsc{FWHM} may overestimate the ejecta velocity; therefore, we adopted the relation proposed by \citet{2024MNRAS.530.4531S_Santamaria}:
\begin{equation}\label{santamaria}
	V_{exp} = \frac{\mathrm{FWHM}}{2\sqrt{\ln 2}}.
\end{equation}

By applying Equation~\ref{santamaria} to the FWHM values, corrected for instrumental broadening, we estimate the minimum and maximum expansion velocities, which in turn allow us to determine the inner and outer radii of the ejected shell. The obtained minimum and maximum velocities are about 790 \kms and 1220 \kms, respectively. Based on these values, the inner and outer radii for each epoch are estimated to be in the range of 1.63 $\times 10^{15}$cm to 7.08 $\times 10^{15}$cm , and 2.52 $\times 10^{15}$cm to 11.07 $\times 10^{15}$cm , respectively (see the values in Table \ref{tab:T2_results}). This estimate aligns well with the findings of \citet{2023RNAAS_277R_Rudy}, who reported an approximate value of 5.2 $\times 10^{15}$ cm based on the expansion velocity derived from VLA measurements and the GAIA distance.

A set of model spectra is generated by systematically varying all previously mentioned input parameters over a broad range with fine increments. The temperature spans from $10^{4.0}$ to $10^6$~K, luminosity from $10^{35}$ to $10^{40}$~\ergs, and ejecta density from $10^{5.0}$ to $10^{10}~\text{cm}^{-3}$, alongside adjustments in elemental abundances. Several trial models are explored iteratively at each epoch until a satisfactory final model is obtained. Initially, a visual inspection is carried out to eliminate models that clearly do not reproduce the observed spectrum. Subsequently, the goodness of fit is assessed by calculating the $\chi^2$ and reduced $\chi^2$ ($\chi^2_{\mathrm{red}}$) values. The uncertainty $\sigma$ is typically in the range of 10\%–30\%, depending on the line strength relative to the continuum and on possible blending with nearby spectral features \citep{Helton10}. Line fluxes are interactively measured by fitting Gaussian profiles using the \textit{splot} task from the \textit{onedspec} package in \textsc{iraf}. A good model is expected to yield $\chi^2 \approx \nu$ \citep{2001MNRAS-Schwarz}, and an acceptable reduced chi-square value, $\chi^2_{\text{red}}$, generally falls between 1 and 2, indicating a reliable fit. The optimal model parameters along with their estimated uncertainties are listed in Table~\ref{tab:T2_results}. These uncertainties are determined by varying one parameter at a time while keeping the others fixed at their best-fit values, continuing until $\chi^2_{\text{red}}$ reaches 2. This method provides an uncertainty estimate approximately equivalent to 3$\sigma$ \citep{2001MNRAS-Schwarz}.

This study involves modeling seven spectra, corresponding to epochs~1 through~7. These spectra were obtained on days 240, 339, 432, 626, 683, 980, and 1051 after the outburst. Except for the spectra observed on days 240, 339, and 432---which cover the wavelength range 3800--7550~\AA---all other modeled spectra span $\sim$ 3700--9000~\AA. However, since no prominent emission features were present redward of 7500~\AA, we restrict our presentation to the 3800---7500~\AA\ range.  The observations were made in two frames: grism 7 (3700 - 7000 \AA) and grism 8 (5200 - 9000  \AA). After applying standard data reduction procedures of \textsc{iraf}, we combined the frames of each corresponding epoch using the \textsc{iraf} task \textit{scombine}. We chose to start modeling after 240 days post-outburst because the nova is very slow and exhibited several rebrightening events. Consequently, photoionization may not have been the dominant process in the early days.

In this study, we found that a one-component model of \textsc{cloudy} is sufficient to meet our needs. Although we ran several two-component models for verification, we did not encounter any critical situations that convinced us to prefer the two-component model over the one-component model. The one-component model allowed us to generate almost all lines within a reasonably acceptable fitting range, and the two-component model did not provide any new emission lines. Therefore, we decided to stick with the one-component model and proceed with further analysis. This might indicate that the ejecta in this nova has less heterogeneity or variation compared to fast novae such as V1674 Her \citep{2024MNRASHabtie}. This could be attributed to the slowness of nova V1405 Cas in evolving, resulting in slower variations in density and other parameters. However, the covering factor, which represents the fraction of $4\pi$ sr subtended by the gas as seen from the central source, shows only slight variations across epochs and remains consistently below unity in all cases (see Table~\ref{tab:T2_results}).

\begin{figure*}
	\centering
	\includegraphics[scale=0.5]{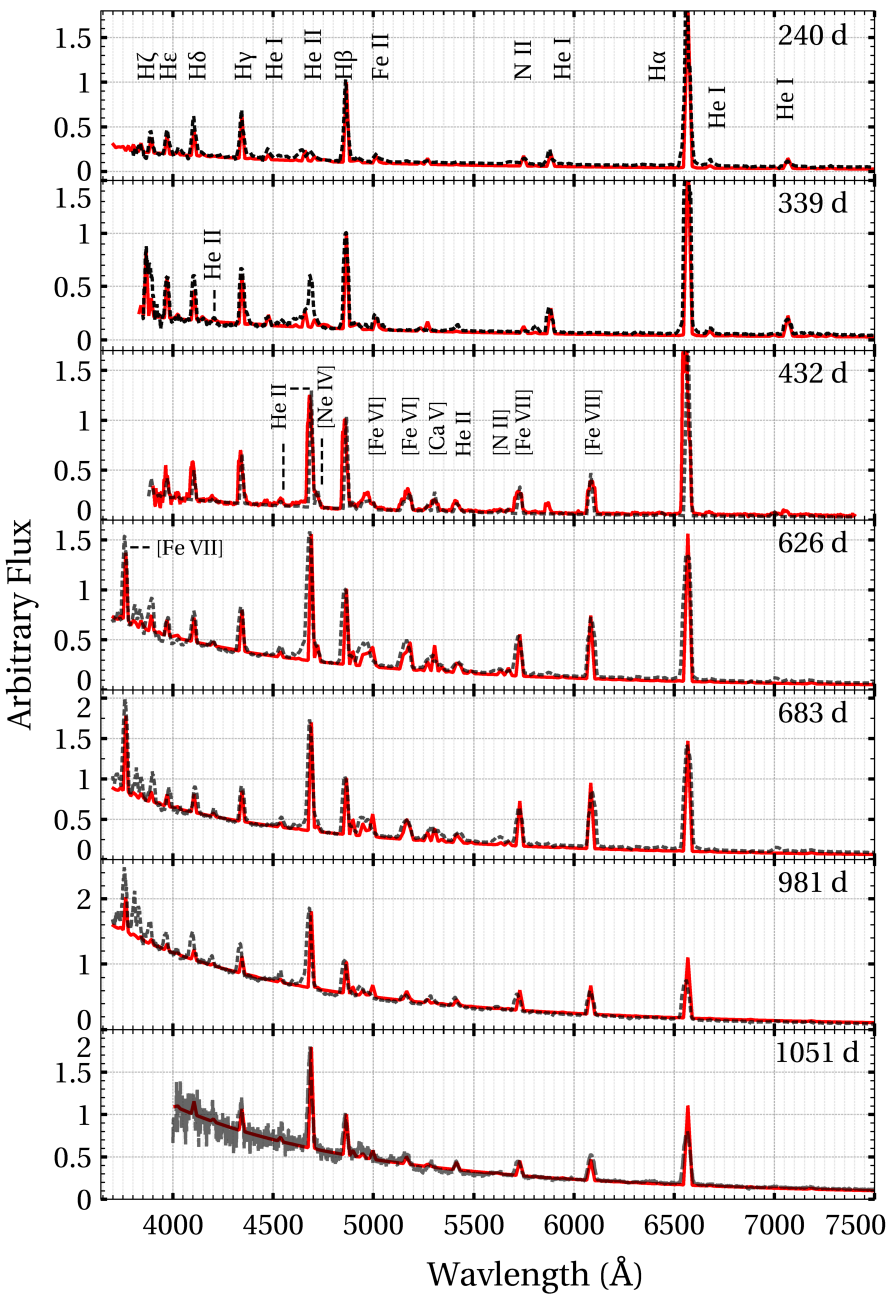} 
	\caption{The best-fitting \textsc{cloudy} synthetic spectra (red solid lines) are overplotted on the observed spectra (black dashed lines) of V1405 Cas, obtained on  day 240 (2021 November 13.39 UT), day 339 (2022 Feb 19.42 UT), day 432 (2022 May 24.22 UT), day 626 (2022 December 3.66 UT), day 683 ( 2023 January 30.58 UT), day 981 (2023 November 23.76 UT), and day 1051 (2024 February 2.56 UT), from top to bottom, respectively. Both the observed and modeled spectra are normalized to H$\beta$, and the observed spectra have been corrected for reddening using E(B–V) = 0.53.}	
	\label{fig:cloudy_best_fitting_model}
\end{figure*}

Fig.~\ref{fig:cloudy_best_fitting_model} displays the best-fitting \textsc{cloudy} model spectra (in red) overlaid on the observed optical spectra (in black) across seven different epochs. A comparison of relative fluxes between the observed lines and those predicted by the best-fitting model during the early phase is provided in Table~\ref{tab:T3_chi_square}, along with the corresponding $\chi^2$ and $\chi^2_{\text{red}}$ values. The $\chi^2$ values are computed using only those emission lines that are present in both the observed and model spectra. Line fluxes for both observations and models are measured using \textsc{iraf}, with multi-component line profiles fitted using multiple Gaussian functions. To minimize the effects of flux calibration uncertainties across epochs, the fluxes of both observed and modelled lines are normalized relative to H$\beta$.

\begin{deluxetable*}{lccccccc}
	\tablecaption{Best-fitting \textsc{cloudy} model parameters and values for nova V1405~Cas.}\label{tab:T2_results}
	\tablewidth{0pt}
	\tabletypesize{\scriptsize}
	\tablehead{
		\colhead{Parameters (units)} & \colhead{Day 240} &\colhead{Day 339}&\colhead{Day 432}& \colhead{Day 626} & \colhead{Day 683} & \colhead{Day 981} & 	\colhead{Day 1051}
	}
	\startdata
	BBT ($\times 10^5$ K)                                       & 0.316$\pm$0.015   &0.347$\pm$0.016&1.202$\pm$0.057& 1.148$\pm$0.054 & 1.259$\pm$0.029 & 1.259$\pm$0.029 & 1.549$\pm$0.029 \\
	L($\times 10^{38}$\ergs)      \dotfill                      & 0.631$\pm 0.077$  &0.794$\pm$0.206&1.000$\pm$0.259& 1.122$\pm 0.080$  & 1.259$\pm 0.090$  & 2.512$\pm 0.118$  & 3.981$\pm 0.188$ \\
	n$_H$ ($\times 10^{6}\text{cm}^{-3}$) \dotfill              & 50.12$\pm 2.362$  &25.12$\pm$3.065&2.512$\pm$0.307& 0.891$\pm 0.021$  & 0.631$\pm 0.031$  & 0.302$\pm 0.014$  & 0.263$\pm 0.006$ \\
	N$_H$ ($\times 10^{21}\text{cm}^{-2}$) \dotfill             & 28.77$\pm$1.36   &19.38$\pm$2.37  & 2.39$\pm$0.29 &1.33$\pm$0.03  & 1.04$\pm$0.05  & 0.71$\pm$0.03 & 0.66$\pm$0.02 \\
	$\alpha$                   \dotfill                         & -2.000            &-2.000&-2.000 & -2.000  & -2.000  & -2.000  & -2.000 \\
	R$_{in}$($\times 10^{15}$cm)$^a$ \dotfill                   & 1.633             &2.371&3.090& 4.276  & 4.667  & 6.699  & 7.081 \\
	R$_{out}$($\times 10^{15}$cm)$^a$  \dotfill                 & 2.518             &3.515&4.467& 6.577 & 7.195  & 10.33  & 10.97 \\
	Filling Factor$^a$          \dotfill                        & 0.100             &0.100&0.100& 0.100   & 0.100   & 0.100   & 0.100  \\
	$\beta$ $^a$                 \dotfill                       & 0.000             &0.000&0.000& 0.000   & 0.000   & 0.000   & 0.000  \\
	Covering Factor              \dotfill                       & 0.850             &0.850&0.850& 0.870   & 0.850   & 0.900   & 0.900  \\
	He/He$_{\sun}$~$^b$  \dotfill         & 1.600$\pm 0.400$  &2.500$\pm$0.500&1.900$\pm$0.300& 2.100$\pm 0.100$   & 2.400$\pm 0.100$   & 3.000$\pm 0.200$   & 3.000$\pm 0.200$  \\
	N/N$_{\sun}$~$^b$    \dotfill         & 1.000$\pm 0.500$  &1.000$\pm 0.500$&1.000$\pm 0.500$& 0.900$\pm 0.300$   & 0.900$\pm 0.500$   & 1.100$\pm 0.700$   & 1.500$\pm 0.600$  \\
	Fe/Fe$_{\sun}$~$^b$  \dotfill       & 0.400$\pm 0.200$  &0.600$\pm$0.300&0.140$\pm$0.040& 0.290$\pm 0.200$&0.350$^{+0.300}_{-0.500}$ & 0.280$\pm 0.300$& 0.600$\pm 0.100$  \\
	Ne/Ne$_{\sun}$~$^b$    \dotfill       & 0.600$\pm 0.300$  &0.600$\pm$0.200&0.600$\pm$0.300& 1.000$\pm 0.200$   & 1.000$\pm 0.700$   & 1.400$\pm 0.300$   & 1.500$\pm 0.200$  \\
	Ca/Ca$_{\sun}$~$^b$   \dotfill      & 0.800$\pm 0.400$  &0.8$\pm$0.200&0.500$\pm$0.200& 0.800$\pm 0.300$   & 1.000$\pm 0.300$   & 0.300$\pm 0.200$   & 0.200$^{+0.200}_{-0.100}$  \\
	M$_{eje}$($\times 10^{-4}M_{\sun}$) \dotfill                & 1.06              &1.705&0.349& 3.27  & 3.11  & 0.47  & 0.50 \\
	Number of lines                     \dotfill                & 15                &16   &15    & 21     & 21     & 20     & 18    \\
	Free parameters                     \dotfill                & 8                 &8     &8    & 8          & 8         & 8          & 8         \\
	DoF                   \dotfill                              & 7                 &8    &7    & 13        & 13         & 12        & 10       \\
	$\chi_{\text{tot}}^2$                 \dotfill              & 12.81             &10.693&8.261& 20.53  & 23.70  & 22.32  & 18.37 \\
	$\chi_{\text{red}}^2$                 \dotfill              & 1.829             &1.337&1.180& 1.579  & 1.823  & 1.860  & 1.837 \\	
	\enddata
	\tablecomments{%
		$^{a}$~This was not a free parameter in the model. 
		$^{b}$~Elemental abundances are expressed on a logarithmic scale relative to hydrogen, based on solar number abundances, with values such as He = $-1.07$, N = $-4.17$, Ne = $-4.07$, Ca = $-5.66$, and Fe = $-4.50$ \citep{2001AIPC..598...23H, 2009ARAA..47..481A}. Full forms of abbreviations: BBT = Blackbody Temperature, L = Luminosity, R$_{\rm in}$ = Inner radius, R$_{\rm out}$ = Outer radius, DoF = Degrees of freedom. %
	}
\end{deluxetable*} 
 
\subsubsection{Temperature and Luminosity}
The parameter values derived from the best-fitting \textsc{Cloudy} models are presented in Table~\ref{tab:T2_results}. These optimal models indicate that the effective temperature of the source increased from $3.16 \times 10^{4}$~K in epoch~1 to $1.55 \times 10^{5}$~K in epoch~7. The largest increase between consecutive epochs is observed between epoch~2 and epoch~3, where the temperature rises by a factor of 3.5. In contrast, the temperature changes between epochs~3~\&~4, ~4~\&~5, 5~\&~6, and 6~\&~7 are relatively modest, with factors of approximately 1.0, 1.1, 1.0 and 1.2, respectively. The significant temperature rise in epoch~3 is primarily attributed to the nova entering its coronal phase on day~371 (see Section~\ref{lp}). This result is consistent with previous reports indicating that the supersoft X-ray source (SSS) turned on around day~300 and continued to brighten, reaching its maximum observed brightness near day~1000 \citep{2021ATel15111....1P, 2024ATel16876....1P}. Correspondingly, the luminosity of the system rose from $6.31 \times 10^{37}$~\ergs~ during epoch 1 to $3.98 \times 10^{38}$~\ergs~ by epoch 7. The effective temperature and luminosity values from our best-fitting \textsc{cloudy} models are reasonably consistent with previously estimated values of $T = 1.6 \times 10^{5}$~K and $L = 9.48 \times 10^{37}$~\ergs~ for day 863 \citep{2023RNAAS_277R_Rudy}.  The apparent rise in temperature and luminosity is most likely due to the gradual thinning of the pseudo-photosphere and expanding ejecta, which became optically thin to soft X-rays between approximately days~100 and 300. As the ejecta progressively dissipate, the optically thick photosphere contracts, and the supersoft X-ray source (SSS) emission becomes visible as deeper layers are revealed, resulting in an apparent increase in the temperature of the underlying WD. This interpretation is supported by the decline in the model-derived column density ($N_{\rm H}$; see Section~\ref{ed}) and is consistent with the expected behaviour during the supersoft X-ray phase \citep[e.g.,][]{1989agnaOsterbrock, 2011ApJ...733...70N}.

\subsubsection{Ejecta Density} \label{ed}
Throughout epochs 1 to 7, the hydrogen density varies from $5.01 \times 10^{7}$ to $2.63 \times 10^{5}$ $\text{cm}^{-3}$. Given the extended timescale under consideration, the expansion phase of this nova outburst is characterized by a gradual decrease in the density of the ejected material. This slow decline results primarily from the low-velocity outward motion of the ejecta, some of which is absorbed by surrounding interstellar matter or gas from the companion star. This absorption adds energy to the nova system, promoting a gradual transition to an optically thin state and further reducing the density. As the material becomes less dense, ionizing radiation from the central WD increasingly penetrates the ejecta, enhancing the ionization level. This process leads to the emergence of emission lines corresponding to higher ionization potentials. Overall, the observed evolution in density and ionization in novae can be attributed to the interplay between slow expansion, material absorption, and the ionizing radiation from the central WD.

The hydrogen column density, $N_\mathrm{H}$, represents the number of hydrogen atoms per unit area along the line of sight and is a key parameter for understanding the optical depth and X-ray transparency of nova ejecta. It can be estimated from the hydrogen number density, $n_\mathrm{H}$, provided by our photoionization models, by integrating along the radial extent of the ejecta \citep{2015PASP..127..266M}:
\[
N_\mathrm{H} = \int_{R_\mathrm{in}}^{R_\mathrm{out}} n_\mathrm{H}(r) \, dr
\]
For a density profile following $n_\mathrm{H}(r) \propto r^{-2}$, appropriate for freely expanding ejecta, the column density simplifies to $N_\mathrm{H} \approx n_\mathrm{H,in} R_\mathrm{in} \left(1 - R_\mathrm{in}/R_\mathrm{out}\right)$, where $n_\mathrm{H,in}$ is the hydrogen number density at the inner radius $R_\mathrm{in}$. Using the values derived from our seven modeled epochs, we obtain column densities decreasing from $N_\mathrm{H} = (2.88 \pm 0.14) \times 10^{22}\,\mathrm{cm^{-2}}$ at epoch~1 to $N_\mathrm{H} = (6.6 \pm 0.2) \times 10^{20}\,\mathrm{cm^{-2}}$ at epoch~7 (see Table~\ref{tab:T2_results}). This monotonic decline is consistent with the continued expansion and dilution of the ejecta, gradually rendering it optically thin and allowing the previously obscured supersoft X-ray source to become visible. 

Using the adopted reddening value of $E(B-V)=0.53$~mag and the empirical relation of \citet{2009MNRAS.400.2050G}, we derived a hydrogen column density of $N_{\rm H}\approx3.63\times10^{21}~\mathrm{cm^{-2}}$. This value is broadly consistent with the X-ray-derived estimate of $N_{\rm H}\approx2.00\times10^{21}~\mathrm{cm^{-2}}$ reported by \citet{2024ATel16876....1P}. Since this estimate is derived solely from the interstellar reddening, it primarily traces the interstellar medium (ISM). Consequently, it is expected to be lower than the modeled column density during the early stages of the outburst, when the total column density includes contributions from both the ISM and the expanding nova ejecta. As the ejecta expand and become progressively more diffuse, the total column density is expected to gradually approach the ISM value. However, the lower $N_{\rm H}$ values derived for the late epochs, particularly epochs~6 and 7, which fall considerably below the estimated interstellar column density, are not physically expected. We suggest that this discrepancy arises from the limitations of the one-dimensional \textsc{CLOUDY} modeling, including the simplified assumptions regarding the ejecta geometry, density distribution, filling factor, and ionization structure.

\subsubsection{Elemental Composition} \label{ec} 
Our photoionization modeling of the ejecta from nova V1405 Cas reveals a composition primarily consisting of He, N, Fe, Ne, and Ca, with noticeable temporal variations in their relative abundances. Across all epochs, helium is significantly enhanced, with a median abundance of  He/He$_\odot \approx 2.4$. In contrast, calcium and iron remain subsolar throughout, with median abundances of Ca/Ca$_\odot \approx 0.63$ and Fe/Fe$_\odot \approx 0.35$,  respectively. The median abundances of nitrogen and neon remain close to their solar values (N/N$_\odot \approx 1.0$ and Ne/Ne$_\odot \approx 1.0$). These median abundances are adopted as representative of  the ejecta composition, while acknowledging that the observed epoch-to-epoch  variations may primarily reflect changes in ionization conditions rather than intrinsic abundance differences.

The persistent overabundance of helium across all epochs suggests significant enrichment, likely resulting from nuclear processing during the outburst. Such an enhancement may also indicate mixing between the accreted hydrogen-rich envelope and the underlying helium layer of the WD, a process commonly referred to as He-mixing \citep{2011arXiv1111.6777G, 2021MNRAS.505.2975G, 2022A&A...660A..53G}. \citet{1998ApJ...494..680J} reported that WDs with masses in the range $0.8$--$1.35\,M_{\odot}$ exhibit varying degrees of mixing—ranging from approximately 25\% to 75\%—between the accreted hydrogen-rich envelope and the outermost layers of the WD core.
 
On the other hand, the persistent underabundance of Fe and Ca suggests possible depletion, likely due to their incorporation into dust grains or a low initial abundance in the accreted material, rather than nuclear processing, which is inefficient for such heavy elements under typical nova conditions \citep{1998PASPGehrz, 1998ApJ...494..680J}. Although there is no strong evidence for dust formation in the case of V1405 Cas \citep{2021ATel14665....1B}, the low initial abundance of these elements is the most likely explanation for their sub-solar levels in the ejecta. 

The abundances of nitrogen and neon do not show statistically significant variations across the different epochs within the uncertainties.  Nitrogen, like helium, is likely produced by dredge-up of material from the underlying WD core, whereas neon is possibly inherited from the donor star, with at most a minor contribution from limited WD mixing. A comprehensive summary of the elemental abundances and their evolution over time is provided in Table~\ref{tab:T2_results}. 

In our photoionization models, no significant enhancement of neon was detected. Of the seven modeled epochs, five exhibited solar or subsolar neon abundances, while only two epochs showed a marginal increase (see Table~\ref{tab:T2_results}). The slight neon enhancement observed in the final two epochs can be naturally explained without invoking an ONe WD. CO WDs inherently contain a non-negligible amount of neon, particularly the isotope \({}^{22}\mathrm{Ne}\), which is produced during the helium-burning phase through the reaction chain $
{}^{14}\mathrm{N}(\alpha,\gamma){}^{18}\mathrm{F}(e^{+}\nu)\,{}^{18}\mathrm{O}(\alpha,\gamma){}^{22}\mathrm{Ne}$ \citep{2022A&A...660A..53G}. As a consequence, \({}^{22}\mathrm{Ne}\) becomes the dominant neon isotope in CO WDs. Therefore, the detection of neon lines or mild neon enrichment in nova ejecta does not uniquely imply an ONe WD; a detailed abundance analysis is required to reliably distinguish between CO and ONe novae.

Based on the results presented in this paper, we find no direct spectroscopic evidence in the optical band to classify this nova as a neon type. Consequently, we question the earlier classifications by \citet{2022ATel15796Munari, 2023ApJTaguchi}, which were primarily based on the presence of strong neon lines and an inferred aluminum overabundance. Notably, the neon abundance derived by \citet{2024AstL...50..317T}, together with the photoionization modeling results presented here, indicates subsolar or near-solar neon abundances. Such low neon levels are not characteristic of typical novae happening on ONeMg WDs and do not allow one to confidently assert that the WD in this system is of the ONeMg type. Furthermore, aluminum enrichment is not unique to novae happening on ONeMg WDs and has also been observed in some CO novae (e.g., V3890~Sgr; \citealt{2020ApJ...895...80O}), while, conversely, certain confirmed novae happening on ONeMg WDs do not exhibit enhanced aluminum abundances (e.g., V723~Cas; \citealt{2018MNRAS.473..355T}).

We also find no convincing evidence for carbon enhancement in the optical spectra. This non-detection does not necessarily imply that carbon is absent from the ejecta, but rather reflects the fact that strong low-excitation carbon transitions are not expected to produce prominent emission in the optical under typical nova ionization and excitation conditions. Carbon diagnostics are generally stronger in the ultraviolet and near-infrared, and their absence in the optical therefore does not rule out a CO WD origin. Further investigation at other wavelength ranges, such as the ultraviolet and near-infrared, is required to better constrain the nova type of V1405~Cas.

\begin{deluxetable*}{llccccccccccccccccccccc} 
	\setlength{\tabcolsep}{1pt}
	\tablecaption{Observed and best-fitting \textsc{cloudy} model line fluxes for V1405 Cas.} 
	\label{tab:T3_chi_square}
	\tabletypesize{\scriptsize}
	\tablehead{
		\colhead{LineID} & \colhead{$\lambda$ (\AA)} & \multicolumn{3}{c}{Day 240} & \multicolumn{3}{c}{Day 339} & \multicolumn{3}{c}{Day 432}  & \multicolumn{3}{c}{Day 626} & \multicolumn{3}{c}{Day 683} & \multicolumn{3}{c}{Day 981} & \multicolumn{3}{c}{Day 1051} \\
		\cline{3-5} \cline{6-8} \cline{9-11} \cline{12-14} \cline{15-17} \cline{18-20} \cline{21-23}
		\colhead{} & \colhead{} &
		\colhead{\shortstack{\\[0.5ex]Observed\\Flux}} & \colhead{\shortstack{\\[0.5ex]Model\\Flux}} & \colhead{$\chi^2$} &
		\colhead{\shortstack{\\[0.5ex]Observed\\Flux}} & \colhead{\shortstack{\\[0.5ex]Model\\Flux}} & \colhead{$\chi^2$} &
		\colhead{\shortstack{\\[0.5ex]Observed\\Flux}} & \colhead{\shortstack{\\[0.5ex]Model\\Flux}} & \colhead{$\chi^2$} &
		\colhead{\shortstack{\\[0.5ex]Observed\\Flux}} & \colhead{\shortstack{\\[0.5ex]Model\\Flux}} & \colhead{$\chi^2$} &
		\colhead{\shortstack{\\[0.5ex]Observed\\Flux}} & \colhead{\shortstack{\\[0.5ex]Model\\Flux}} & \colhead{$\chi^2$} &
		\colhead{\shortstack{\\[0.5ex]Observed\\Flux}} & \colhead{\shortstack{\\[0.5ex]Model\\Flux}} & \colhead{$\chi^2$} &
		\colhead{\shortstack{\\[0.5ex]Observed\\Flux}} & \colhead{\shortstack{\\[0.5ex]Model\\Flux}} & \colhead{$\chi^2$}
	}
	\startdata
	H$\zeta$     	&3889&0.26&0.16&0.45&0.48&0.18&3.90&-&-&-&0.74&0.68&0.12&0.37&0.28&0.19&0.47& 0.23& 1.61&-&-&- \\
	H$\epsilon$  	&3970&0.34&0.20&0.88&0.37&0.37&0.00&0.28&0.17&1.77&0.28&0.19&0.32&0.29&0.22&0.11&0.39&0.21&0.91&-&-&- \\
	H$\delta$    	&4101&0.44&0.27&1.23&0.39&0.28&0.50&0.39&0.27&2.36&0.42&0.29&0.61&0.38&0.30&0.16&0.76&0.33&5.02&0.12&0.09&0.63 \\
	He \textsc{ii}  &4200&-&-&-&0.04&0.03&0.00&0.03&0.03&0.00&0.25&0.12&0.58&0.09&0.10&0.01&0.28&0.12& 0.70&0.10&0.09&0.00 \\
	H$\gamma$    	&4340&0.57&0.49&0.32&0.60&0.48&0.65&0.56&0.49&0.76&0.62&0.49&0.60&0.53&0.53&0.00&0.72&0.47&1.72&0.40&0.45&0.08 \\
	He \textsc{i}   &4471&0.13&0.14&0.00&0.09&0.10&0.01&-&-&-&-&-&-&-&-&-&-&-&-&-&-&- \\
	He \textsc{ii}  &4542&-&-&-&-&-&-&1.33&0.73&0.01&0.19&0.11&0.22&0.07&0.10&0.02&0.14&0.14&0.00&0.12&0.13&0.01 \\
	N \textsc{iii}	&4640&-&-&-&-&-&-&-&-&-&0.51&0.35&0.83&-&-&-&-&-&-&-&-&- \\
	He \textsc{ii}	&4686&0.29&0.12&1.18&0.46&0.17&3.80&1.22&1.24&0.07&1.41&1.48&0.16&1.60&1.76&0.62&2.22&2.44&1.38&2.22&1.93&2.93 \\
	He \textsc{i}  	&4713&-&-&-&-&-&-&-&-&-&0.08&0.22&0.69&0.13&0.39&1.60&-&-&-&-&-&- \\
	H$\beta$ 		&4861&1.00&1.00&0.00&1.00&1.00&0.00&1.00&1.00&0.00&1.00&1.00&0.00&1.00&1.00&0.00&1.00&1.00&0.00& 1.00&1.00&0.00\\
	He \textsc{i}  	&4922&0.12&0.21&0.37&0.06&0.07&0.01&-&-&-&-&-&-&-&-&-&-&-&-&-&-&- \\
	{[Fe \textsc{vii}]}&4942&-&-&-&-&-&-&0.20&0.21&0.01&0.69&1.00&3.20&0.67&1.18&6.42&0.92&1.00&0.22&0.91&0.54&4.93 \\
	He \textsc{i}  	&5016&0.20&0.12&0.31&0.14&0.14&0.00&-&-&-&-&-&-&-&-&-&-&-&-&-&-&- \\
	{[Fe \textsc{vi}]}&5159&-&-&-&-&-&-&0.39&0.42&0.11&0.54&0.97&6.40&0.44&1.17&1.31&0.43&0.70&2.08&0.62&0.39&1.89 \\
	{[Ca \textsc{v}]}&5281&-&-&-&-&-&-&0.21&0.07&2.82&0.41&0.38&0.03&0.42&0.37&0.05&0.42&0.16&1.78&0.42&0.14&2.57 \\
	He \textsc{ii}	&5412&0.04&0.02&0.01&0.04&0.02&0.03&0.12&0.15&0.08&0.19&0.32&0.60&0.19&0.36&0.72&0.30&0.31&0.00&0.28&0.20&0.22 \\
	{[Ca \textsc{vii}]}&5619&-&-&-&-&-&-&-&-&-&0.10&0.10& 0.00&.15&0.13&0.01&0.14&0.08&0.08&0.20&0.05&0.72 \\
	{[Fe \textsc{vi}]}&5714&-&-&-&-&-&-&-&-&-&0.56&0.55&0.01&0.41&0.51&0.25&0.35&0.41& 0.13&0.42&0.35&0.15 \\
	{[N \textsc{ii}]}&5746&0.12&0.02&0.47&0.02&0.10&0.25&-&-&-&-&-&-&0.38&0.43&0.07&0.15&0.29&0.55&0.18&0.32&0.67 \\
	He \textsc{i} 	&5876&0.19&0.21&0.01&0.28&0.33&0.12&-&-&-&-&-&-&-&-&-&-&-&-&-&-&- \\
	{[Fe \textsc{vii}]}&6086&-&-&-&-&-&-&0.54&0.56&0.08&0.91&0.90&0.01&1.10&1.17& 0.12&0.85&0.73&0.39&0.84&0.82&0.01 \\
	H$\alpha$  		&6562&2.97&2.56&7.42&2.84&2.71&0.76&2.69&2.69&0.00&2.20&2.42&1.62&2.36&2.44&0.19&1.50&1.96&5.73&1.79&2.10&3.47\\
	He \textsc{i} 	&6678&0.11&0.06&0.10&0.09&0.07&0.01&0.04&0.03&0.01&0.07&0.07&0.00&0.06&0.03&0.03&0.04&0.03&0.00&0.08&0.03&0.07 \\
	He \textsc{i} 	&7065&0.16&0.20&0.06&0.20&0.32&0.65&0.06&0.03&0.18&-&-&-&-&-&-&-&-&-&-&-&- \\
	{[Fe \textsc{ii}]}&7155&-&-&-&-&-&-&-&-&-&0.07&0.06&0.00&0.04&0.04&0.00&0.055&0.04&0.02&0.10&0.09& 0.00 \\
	He \textsc{ii}	&8237&-&-&-&-&-&-&-&-&-&0.05&0.04& 0.00&0.08&0.05&0.03&0.07&0.05&0.01&0.10&0.08&0.01 \\	
	\enddata
	\tablecomments{%
	The H$\zeta$ and H$\eta$ features may be blended with the expected [Ne III] 3869 \AA~ and 3968 \AA~ emission, respectively. Because the Balmer and neon components cannot be reliably separated in the present spectra, the presence of [Ne III] cannot be confirmed directly.
	}
\end{deluxetable*}

\subsubsection{Ejected Mass Calculation} \label{em}
The ejecta mass in the expanding shell (\(M_{eje}\)) within a spherical volume defined by inner and outer radii (\(\mathrm{r_{in}}\) and \(\mathrm{r_{out}}\)) can be calculated in units of \(M_{\odot}\) by integrating the number density and filling factor functions given in Equation~\ref{ndff} over the radial range of the spherical volume, as expressed by:
\begin{equation}\label{mej}
	M_{eje} = 4\pi \times m_H \times n_H \times \text{cf} \times \text{ff} \times \mathrm{r^2_{in}} \times (\mathrm{r_{out}} - \mathrm{r_{in}}),
\end{equation}
where \(m_H\) is the mass of a single hydrogen atom, equal to \(1.67 \times 10^{-24}\)g; 'cf' is the covering factor; \(n_H\) is the hydrogen number density; and 'ff' is the filling factor. Then using the parameter values listed in Table~\ref{tab:T2_results}, we computed the ejecta mass of the expanding shell at seven epochs. The derived masses for epochs~1--7 are $1.06\times10^{-4}\,M_{\odot}$, $1.71\times10^{-4}\,M_{\odot}$, $0.35\times10^{-4}\,M_{\odot}$, $3.27\times10^{-4}\,M_{\odot}$, $3.11\times10^{-4}\,M_{\odot}$, $0.47\times10^{-4}\,M_{\odot}$, and $0.50 \times10^{-4}\,M_{\odot}$, respectively, with a mean value of $\sim1.5\times10^{-4}\,M_{\odot}$. The relatively higher masses obtained for epochs~4 and 5 are not expected to represent actual changes in the total ejecta mass, but rather reflect the sensitivity of the derived mass to the adopted density distribution, filling factor, and geometry at each epoch.

As an independent estimate, we also derived the ejecta mass using the helium abundance factor $Y$ \citep{1993AJ....106.2408S}. The resulting masses are $0.79\times10^{-4}\,M_{\odot}$, $0.63\times10^{-4}\,M_{\odot}$, $0.73\times10^{-4}\,M_{\odot}$, $0.69\times10^{-4}\,M_{\odot}$, $0.65\times10^{-4}\,M_{\odot}$, $0.58\times10^{-4}\,M_{\odot}$, $0.58\times10^{-4}\,M_{\odot}$, with an average value of $0.66\times10^{-4}\,M_{\odot}$. Since the two methods provide independent estimates based on different assumptions, we adopt their average as a more
representative ejecta mass, yielding $M_{\rm ej}\approx1.1\times10^{-4}\,M_{\odot}$.

The high ejecta mass inferred from our photoionization modeling suggests a relatively low-mass WD, disfavouring an ONeMg nova interpretation. Both theoretical models and observational studies indicate that novae erupting on lower-mass CO WDs accumulate more material prior to ignition and consequently eject larger envelope masses during outburst \citep[e.g.,][]{1998MNRAS.296..502S, 2005ApJ...623..398Y}. In contrast, novae originating from more massive ONeMg WDs are expected to produce smaller ejecta masses due to stronger surface gravity and earlier thermonuclear ignition \citep{2022MNRAS.516.1008L, 1998PASPGehrz}. Ejecta masses of order $\sim10^{-4}\,M_\odot$ are therefore more consistent with slow CO novae than with typical ONeMg systems.

\section{Conclusion}\label{sec7_v1405cas}
We conducted a detailed investigation of the spectroscopic and photometric evolution of V1405~Cas, one of the slowest  nova ever recorded. Using the photoionization code \textsc{Cloudy}, we modeled the observed spectra to derive key physical and chemical parameters of the system. The main findings of our analysis are summarized below:
\begin{enumerate}
	\item Based on the photometric data, we determined the decline times in the $V$ band to be $t_2 \approx 165$ days and $t_3 \approx 175$ days.  The masses of the WD and the secondary star are calculated to be $\sim$ $0.7\~M_{\odot}$ and $0.43\~M_{\odot}$, respectively. The orbital separation between the primary and secondary stars is found to be $a \approx 1.01 \times 10^{11}~\text{cm}$.

	\item The study suggests that the nova entered the coronal phase by day~+371; however, weak nebular features were already present at earlier epochs (day~+339) and remained suppressed by the dense ejecta. 

	\item The gradual weakening of the He\,\textsc{I} lines, particularly the near disappearance of He\,\textsc{I}\,5876~\AA\ by day~+683, together with the differing evolution of singlet and triplet transitions, indicates a progressive decline in optical depth as the nova ejecta expanded during the late decline phase.
	
	\item During the nebular phase, when the assumptions of Case B are appropriate, the observed hydrogen line ratios are consistent with Case B recombination.
	
	\item The photoionization modeling provides estimates of the source temperature, luminosity, ejecta density, and elemental abundances. Our \textsc{Cloudy} modeling indicates no significant neon overabundance, with neon abundances remaining broadly consistent with solar values across all analyzed epochs. In contrast, helium shows a persistent enhancement, while iron and calcium remain below solar abundances. The mean ejected mass measured in the seven epochs is $\sim1.10\times~10^{-4}~M_{\odot}$. Such a high mean ejected mass is usually expected from a low mass WD. 
	
	\item Based on the analysis of optical photometric and spectroscopic data, supported by detailed photoionization modeling, we find no compelling evidence to classify the host WD as an ONeMg type. A definitive classification will require observations at other wavelengths, particularly in the near-infrared and ultraviolet.
\end{enumerate}

\begin{acknowledgments}
We sincerely thank the anonymous referee for the exceptionally thorough and constructive review, which has significantly improved and substantially reshaped the manuscript. We acknowledge the Debre Berhan University, S. N. Bose National Centre for Basic Science (SNBNCBS) and The World Academy of Science (TWAS) for their funding support.  We thank the staff of IAO, Hanle, CREST, and Hosakote, who made these observations possible. The facilities at IAO and CREST are operated by the Indian Institute of Astrophysics, Bangalore. We are grateful to F. Teyssier for coordinating the Astronomical Rings for Access to Spectroscopy (ARAS) Eruptive Stars Section. Equally, we acknowledge Subhajit Kar, David Boyd, Forrest Sims, and Pavol A. Dubovsky, for their valuable spectroscopic observations. Furthermore, we recognize and value the American Association of Variable Stars Observers (AAAVSO) and \textsc{Vstar} databases  for providing open access to photometric data. This work was also supported by the project APVV-20-0148 From Interacting Binaries to Exoplanets.	
\end{acknowledgments}

\appendix

\section{Measured Values of FWHM, Flux, and EW}
	Table~\ref{tab:FWHM01_v1405} presents the measured values of flux (\(\mathrm{erg\,s^{-1}\,cm^{-2}}\)), full width at half maximum (FWHM, in~\kms), and equivalent width (EW, in~\AA) for the most prominent emission lines across about 21 epochs: ranging from days +2 to 1051. 

\begin{longtable}{lccc|ccc}
		\caption{De-reddened flux, FWHM and EW of most prominent emission lines in the spectrum from Day 1.918 to Day 1051.
		} \\
		\label{tab:FWHM01_v1405}\\
		\hline
\multirow{2}{*}{Days}	& Flux (erg\,cm$^{-2}$\,s$^{-1}$) & FWHM (km\,s$^{-1}$) & EW (\AA) 
& Flux (erg\,cm$^{-2}$\,s$^{-1}$) & FWHM (km\,s$^{-1}$) & EW (\AA) \\
\cline{2-4} \cline{5-7}
& \multicolumn{3}{c}{H$\alpha$} & \multicolumn{3}{c}{H$\beta$} \\	
\hline
2  &1.26E--09 &1193.97  &122.1  &6.15E--10  &1136.19 &41.36 \\
24 &1.54E--09 &612.72  &219.9   &5.75E--10  &596.74  &45.31\\
92 &1.48E--09 &1041.00 &598.00    &4.13E--09  &1063.00 &115.10  \\
170&3.45E--09 &1412.00 &448.8   &1.42E--09  &1253.00 &81.18  \\
240&4.18E--09 &1264.38 &792.50  &1.37E--09  &1399.70 &167.30 \\
284&1.23E--09 &1460.00 &840.40  &5.07E--10  &1558.00 &215.80  \\
303&1.17E--09 &1562.87 &867.60  &4.85E--10  &1653.14 &185.50 \\
338&9.45E--10 &1475.52 &1276.00 &3.31E--10  &1512.53 &245.40  \\
371&5.55E--10 &1481.06 &1097.00 &2.04E--10  &1514.08 &233.30  \\
418&4.33E--10 &1428.56 &902.40  &1.83E--10  &1504.47 &188.20  \\
431&4.08E--10 &1431.53 &1071.00 &1.78E--10  &1493.47 &203.50  \\
453&2.56E--10 &1593.52 &730.70  &1.05E--10  &1533.48 &143.50  \\
480&3.10E--10 &1619.86 &722.60  &1.13E--10  &1657.21 &139.90  \\
494&3.08E--10 &1578.50 &795.80  &1.08E--10  &1626.20 &144.60  \\
523&2.13E--10 &1591.90 &1034.00 &6.25E--11  &1697.90 &187.00 \\
600&1.04E--10 &1638.19 &611.90  &4.19E--11  &1672.12 &111.11  \\
626&5.38E--11 &1730.90 &502.00  &2.41E--11  &1815.40 &94.40 \\
683&2.32E--11 &1669.90 &369.70  &1.03E--11  &1802.80 &68.53 \\
874&7.38E--12 &1630.50 &185.80  &3.27E--12  &1691.00 &38.28 \\
981&3.25E--12 &1626.40 &151.70  &2.16E--12  &1655.50 &28.80 \\
1051&5.75E--13&1568.60 &137.3 0 &2.50E--13  &1598.80 &19.08 \\
\hline
& \multicolumn{3}{c}{H$\gamma$} & \multicolumn{3}{c}{H$\delta$} \\
\hline
2   &3.89E--10 &1075.58  &18.62  &2.54E--10  &1026.09   &10.19 \\
24  &2.98E--10 &577.44   &15.84  &1.93E--10  &823.85   &9.50  \\
92  &3.38E--10 &1144.00  &48.74  &1.58E--10  &1176.00  &30.10 \\
170 &1.21E-09  &1315.00  &40.26  &9.60E--10   &1454.00  &26.27  \\
240 &8.10E--10 &1602.31  &73.24  &6.23E--10  &1576.79  &44.85 \\
284 &3.99E--10 &1787.00  &99.99  &3.57E--10  &1790.00  &55.51  \\
303 &3.44E--10 &1953.03  &92.20  &3.65E--10  &2113.84  &72.10  \\
338 &2.02E--10 & 1804.24 &118.80 &1.36E--10  &1500.49  &62.19  \\
371 &1.22E-10  &1652.70  &100.30 &7.5E--11   &1527.79  &48.85  \\
418 &1.13E--10 &1710.67  &79.55  &6.61E--11  &1535.60  &39.15  \\
431 &9.93E--11 &1643.16  &81.82  &6.92E--11  &1569.58  &45.79  \\
453 &5.40E--11 &1708.50  &60.08  &2.90E--11  &1518.89  &29.60  \\
480 &6.53E--11 &1722.15  &57.10  &3.93E--11  &1625.75  &28.63  \\
494 &6.10E--11 &1754.00  &61.50  &3.73E--11  &1718.50  &32.45  \\
523 &2.80E--11 &1787.60  &77.00  &1.63E--11  &1744.40  &39.92 \\
600 &2.61E--11 &1788.50  &53.71  &1.66E--11  &1753.80  &25.49  \\
626 &1.34E--11 &1750.40  &37.47  &9.36E--12  &1909.70  &21.66 \\
683 &5.63E--12 &1728.70  &25.60  &4.06E--12  &1796.10  &14.96 \\
874 &1.90E--12 &1517.30  &16.66  &1.47E--12  &1609.80  &11.00 \\
981 &1.60E--12 &1559.40  &13.43  &1.57E--12  &1880.70  &10.49 \\
1051&1.18E--13 &1188.20  &9.87   &7.27E--14  &1132.90  &5.35 \\
\hline
& \multicolumn{3}{c}{He I 5876} & \multicolumn{3}{c}{He II 4686} \\
\hline
2  & 2.22E--10 &661.16    &18.40  &    -     &    -   & - \\
24 & 8.37E--11 &578.78   &9.29   &    -     &    -    & -  \\
92 &2.44E--10  &901.00   &8.87   &    -     &    -    & -  \\
170&2.36E--10  &966.00   &23.16  &3.28E--11 &932.00   &1.55  \\
240&2.59E--10  &1170.20  &43.49  &1.13E--10 &1270.50  &9.85 \\
284&1.43E--10  &1408.00  &77.73  &5.59E--10 &2626.00  &180.50  \\
303&1.23E--10  &1436.30  &65.31  &4.60E--10 &2664.33  &162.10  \\
338&9.22E--11  &1386.90  &92.45  &2.26E--10 &2477.19  &149.20  \\
371&2.54E--11  &1359.86  &46.46  &2.56E--10 &1676.45  &275.50  \\
418&2.14E--11  &1179.45  &35.22  &2.65E--10 &1845.83  &252.50  \\
431&1.75E--11  &1123.19  &30.98  &2.42E--10 &1710.01  &248.40  \\
453&9.90E--12  &1236.02  &21.49  &1.55E--10 &1782.13  &201.10  \\
480&1.01E--11  &1271.73  &17.93  &1.69E--10 &1860.64  &181.90  \\
494&9.76E--12  &1203.00  &18.24  &1.50E--10 &1897.60  &179.30 \\
523&6.24E--12  &1393.70  &23.12  &8.04E--11 &1903.50  &230.00\\
600&1.61E--12  &1431.91  &7.40   &6.37E--11 &1728.22  &154.00  \\
626&8.93E--13  &1374.40  &5.82   &4.18E--11 &1987.00  &143.50 \\
683&   -       & -       & -     &1.80E--11 &1843.50  &103.40 \\
874&   -       & -       & -     &6.41E--12 &1692.00  &65.86 \\
981&   -       & -       & -     &5.05E--12 &1797.60  &57.98 \\
1051&  -       & -       & -     &4.83E--13 &1477.70  &45.40 \\
\hline
& \multicolumn{3}{c}{He I 6678} & \multicolumn{3}{c}{He II 7065} \\
\hline
2  &1.26E--10 &722.96    &12.69   &1.30E--10 &761.46  &15.05  \\
24 &2.65E--11 &449.15    &3.76    &2.93E--11 &552.95  &4.82  \\
92 &1.16E--10 &1086.90   &-5.19   &1.02E--10 &1140.34 &5.01  \\
170&1.19E--10 &1167.55   &16.62   &1.04E--10 &1171.90 &15.89  \\
240&1.23E--10  &1311.59  &25.38   &1.14E--10 &1317.37 &27.61  \\
284&4.08E-11   &1382.40  &28.14  &1.02E--10 &1591.43  &79.38  \\
303&3.99E--11  &1621.06  &29.04  &9.50E--11 &1674.54  &76.59  \\
338&2.83E--11  &1533.39  &38.28  &7.05E--11 &1530.50  &98.62  \\
371&7.40E--12  &1527.81  &18.10  &1.48E--11 &1415.14  &36.22  \\
418&6.28E--11  &1379.98  &14.01  &1.01E--11 &1086.08  &23.89  \\
431&8.49E--12  &1640.95  &21.44  &8.30E--12 &1067.58  &19.09  \\
453&3.07E--12  &1398.74  &9.18   &5.36E--12 &1563.85  &17.65  \\
480&4.09E--12  &1421.33  &9.76   &4.75E--12 &1196.39  &12.04  \\
494&4.44E--12  &1447.99  &12.13  &7.89E--11 &1481.04  &24.93  \\
523&4.64E--12  &1559.53  &13.81  &8.44E--12 &1677.13  &29.38  \\
600&6.60E--13  &1290.94  &4.33   &9.24E--13 &1579.07  &7.08  \\
626&4.76E--13  &1424.71  &5.15   &4.95E--13 &1643.88  &6.08 \\
683&2.06E--13  &1299.45  &3.56   &1.45E--13 &1546.06  &2.84 \\
874&7.43E--14  &1068.68  &2.23   &2.21E--14 &881.19   &0.74 \\
981&3.50E--14  &1117.81  &1.84   &7.09E--15 &846.76   &0.44  \\
1051& -        &       - &  -    &    -     &    -    & - \\
\hline
& \multicolumn{3}{c}{[Fe VII] 5721} & \multicolumn{3}{c}{[Fe VII] 6086} \\
\hline
2 -- 338 & - & - &  & - & - & - \\
371&4.63E--11 &1701.44  &75.30  &7.75E--11  &1611.79  &162.10  \\
418&3.15E--11 &1768.38  &46.21  &5.23E--11  &1705.16  &99.34  \\
431&5.92E--11 &1756.01  &95.47  &9.65E--11  &1697.91  &197.40  \\
453&3.21E--11 &1641.54  &60.72  &5.31E--11  &1580.31  &124.10  \\
480&2.89E--11 &1700.10  &43.52  &5.29E--11  &1664.77  &107.70  \\
494&1.76E--11 &1678.60  &28.69  &3.56E--11  &1693.70  &77.90  \\
523&2.82E--11 &1853.50  &97.22  &4.58E--11  &1713.00  &185.70  \\
600&2.00E--11 &1626.57  &77.32  &3.50E--11  &1724.34  &176.00  \\
626&1.37E--11 &1712.30  &75.03  &2.27E--11  &1699.00  &161.20  \\
683&7.17E--12 &1682.70  &75.12  &1.12E--11  &1636.60  &143.70  \\
874&1.36E--12 &1691.00  &38.08  &3.38E--12  &1697.00  &74.78  \\
981&1.13E--12 &1528.00  &28.02  &1.81E--12  &1598.90  &60.86  \\
1051&1.62E--13 &1498.50 &25.72  &2.79E--13  &1565.80  &52.99  \\
\hline
\end{longtable}
\bibliography{sample7new}{}
\bibliographystyle{aasjournal}

\end{document}